\documentclass[10pt]{article}%
\usepackage[a4paper,top=2.6cm,bottom=2.6cm,left=2.9cm,right=2.9cm]{geometry} %
\usepackage{amsmath}%
\usepackage{amsthm}%
\usepackage{amsfonts}%
\usepackage{amssymb}%
\usepackage{mathrsfs}
\usepackage{braket}					
\usepackage{mathtools}
\usepackage{enumitem}
\usepackage{xcolor}%
\usepackage{graphicx}%
\usepackage{bm}
\usepackage{tikz}%
\usetikzlibrary{calc,positioning,automata}%
\usetikzlibrary{cd}%
\usepackage{float}%
\usepackage{etoolbox}
\usepackage{changepage}
\usepackage{upgreek}
\usepackage{nccmath}    
\usepackage{natbib}          
\usepackage[english]{babel}
\usepackage[utf8]{inputenc}
\usepackage[T1]{fontenc}
\usepackage{csquotes} 
\usepackage[left = `,%
right = ',%
leftsub = ``,%
rightsub = '']{dirtytalk} 
\MakeRobust{\say}
\newcommand{\Say}{\enquote}
\usepackage{epigraph}
\makeatletter
\patchcmd{\epigraph}{\@epitext{#1}}{\itshape\@epitext{#1}}{}{}
\makeatother
\usepackage{titlesec}
\titleformat{\section}
  {\normalfont\sc\centering}
  {\thesection}{1em}{}
\titleformat{\subsection}
  {\normalfont\centering}
  {\normalfont\sc\centering\thesubsection}{1em}{}{}
\titleformat{\subsubsection}
  {\normalfont\slshape\centering}
  {\normalfont\sc\centering\thesubsubsection}{1em}{}{}
\titleformat{\paragraph}[runin]
{\normalfont\slshape\normalsize}{\theparagraph}{1em}{}[.]
\makeatletter
\g@addto@macro\@floatboxreset\centering
\makeatother
\makeatletter
\def\bbordermatrix#1{\begingroup \m@th
  \@tempdima 4.75\p@
  \setbox\z@\vbox{%
    \def\cr{\crcr\noalign{\kern2\p@\global\let\cr\endline}}%
    \ialign{$##$\hfil\kern2\p@\kern\@tempdima&\thinspace\hfil$##$\hfil
      &&\quad\hfil$##$\hfil\crcr
      \omit\strut\hfil\crcr\noalign{\kern-\baselineskip}%
      #1\crcr\omit\strut\cr}}%
  \setbox\tw@\vbox{\unvcopy\z@\global\setbox\@ne\lastbox}%
  \setbox\tw@\hbox{\unhbox\@ne\unskip\global\setbox\@ne\lastbox}%
  \setbox\tw@\hbox{$\kern\wd\@ne\kern-\@tempdima\left[\kern-\wd\@ne
    \global\setbox\@ne\vbox{\box\@ne\kern2\p@}%
    \vcenter{\kern-\ht\@ne\unvbox\z@\kern-\baselineskip}\,\right]$}%
  \null\;\vbox{\kern\ht\@ne\box\tw@}\endgroup}
\makeatother
\makeatletter
\newcommand{\mathleft}{\@fleqntrue}
\newcommand{\mathcenter}{\@fleqnfalse}
\makeatother
\DeclareMathOperator{\marg}{marg}
\DeclareMathOperator{\supp}{supp}
\DeclareMathOperator{\proj}{proj}
\DeclareMathOperator{\Bel}{\mbb{B}}
\DeclareMathOperator{\CB}{\mbb{CB}}

\makeatletter
\def\th@plain{%
  \thm@notefont{}
  \itshape 
}
\def\th@definition{%
  \thm@notefont{}
  \normalfont 
}
\makeatother
\numberwithin{equation}{section}
\usepackage{stmaryrd}
\newcommand{\evaluation}[2][]{\ensuremath{\llbracket #2\rrbracket_{#1}}}
\SetSymbolFont{stmry}{bold}{U}{stmry}{m}{n}
\DeclareMathAlphabet{\mathsfit}{\encodingdefault}{\sfdefault}{m}{sl}
\newcommand{\tens}[1]{\mathsfit{#1}}
\DeclareMathAlphabet{\mathbbsf}{U}{bbold}{m}{n}
\newcommand{\mcal}{\mathcal}
\newcommand{\mscr}{\mathscr}

\newcommand{\mbb}{\mathbb}
\newcommand{\mfrak}{\mathfrak}
\newcommand{\Naturals}{\mathbb{N}}

\newcommand{\la}{\langle}
\newcommand{\ra}{\rangle}
\newcommand{\invlim}{\underset{\leftarrow}{\lim}\,}

\newcommand{\CompactFam}{\mathcal{K}}
\newcommand{\Agents}{I}
\newcommand{\Ann}{a}
\newcommand{\Bob}{b}
\newcommand{\ArbSet}{X}

\newcommand{\SNatures}{\Theta}
\newcommand{\snature}{\theta}
\newcommand{\rain}{\mathrm{r}}
\newcommand{\sun}{\mathrm{n}}
\newcommand{\weather}{\mathrm{w}}
\newcommand{\States}{\Omega}
\newcommand{\state}{\omega}
\newcommand{\cprior}{\pi}
\newcommand{\THierSt}{\mscr{T}^{*}}
\newcommand{\Hiers}{T}
\newcommand{\UStates}{\States^{*}}
\newcommand{\UTypes}{T^{*}}
\newcommand{\typehom}{\tau^{*}}
\newcommand{\belhom}{\beta^{*}}

\newcommand{\hierfun}{h}
\newcommand{\projstfun}{\proj}
\newcommand{\TypeSt}{\mscr{T}}
\newcommand{\belfun}{\beta}

\newcommand{\Types}{T}
\newcommand{\type}{t}
\newcommand{\Closure}{\mcal{C}}
\newcommand{\ac}{\mathsf{ac}}
\newcommand{\ADTypeSt}{\mfrak{T}}
\newcommand{\ADTypes}{\tens{T}}
\newcommand{\ADtype}{\tens{t}}
\newcommand{\ADStates}{\Upomega}

\newcommand{\ADbelfun}{\bm{\beta}}
\newcommand{\minADTypeSt}{\widehat{\mfrak{T}}}
\newcommand{\minADTypes}{\widehat{\tens{T}}}
\newcommand{\minADStates}{\widehat{\Upomega}}
\newcommand{\minADbelfun}{\widehat{\bm{\beta}}}
\newcommand{\acADbelfun}{\bm{\alpha}}
\newcommand{\FamADTypeSt}{\bm{\mcal{T}}}
\newcommand{\ADBel}{\pmb{\mathbbsf{B}}}

\newcommand{\ADCB}[1][]{\pmb{\mathbbsf{CB}}^{#1}}
\newcommand{\Event}{E}
\newcommand{\ADEvent}{\tens{E}}
\def\SetVertic{\egroup\;\hspace{-0.1cm}\middle|\hspace{-0.1cm}\;\bgroup}
{\catcode`\|=\active
  \xdef\Sets{\protect\expandafter\noexpand\csname Sets \endcsname}
  \expandafter\gdef\csname Sets \endcsname#1%
  	{\hspace{-0.05cm}\left\{\hspace{-0.08cm}%
     		\:{
     			\mathcode`\|32768\let|\SetVertic
     		#1}
     		\:\hspace{-0.08cm}\right\}}
}

{\catcode`\|=\active
  \xdef\Round{\protect\expandafter\noexpand\csname Round \endcsname}
  \expandafter\gdef\csname Round \endcsname#1{\left(%
     \:{
     \mathcode`\|32768\let|\SetVertic
     #1}\:\right)}
}

{\catcode`\|=\active
  \xdef\Rounds{\protect\expandafter\noexpand\csname Rounds \endcsname}
  \expandafter\gdef\csname Rounds \endcsname#1%
  	{\hspace{-0.05cm}\left(\hspace{-0.08cm}%
     		\:{
     			\mathcode`\|32768\let|\SetVertic
     		#1}
     		\:\hspace{-0.08cm}\right)}
}

{\catcode`\|=\active
  \xdef\Square{\protect\expandafter\noexpand\csname Square \endcsname}
  \expandafter\gdef\csname Square \endcsname#1{\left[%
     \:{
     \mathcode`\|32768\let|\SetVertic
     #1}\:\right]}
}

{\catcode`\|=\active
  \xdef\Squares{\protect\expandafter\noexpand\csname Squares \endcsname}
  \expandafter\gdef\csname Squares \endcsname#1%
  	{\hspace{-0.05cm}\left[\hspace{-0.08cm}%
     		\:{
     			\mathcode`\|32768\let|\SetVertic
     		#1}
     		\:\hspace{-0.08cm}\right]}
}

{\catcode`\|=\active
  \xdef\Angle{\protect\expandafter\noexpand\csname Angle \endcsname}
  \expandafter\gdef\csname Angle \endcsname#1{\left\langle%
     \:{
     \mathcode`\|32768\let|\SetVertic
     #1}\:\right\rangle}
}

{\catcode`\|=\active
  \xdef\Angles{\protect\expandafter\noexpand\csname Angles \endcsname}
  \expandafter\gdef\csname Angles \endcsname#1%
  	{\hspace{-0.05cm}\left\langle\hspace{-0.08cm}%
     		\:{
     			\mathcode`\|32768\let|\SetVertic
     		#1}
     		\:\hspace{-0.08cm}\right\rangle}
}
\usepackage{varioref}
\usepackage[pagebackref,hypertexnames=false]{hyperref}
\usepackage[noabbrev,nameinlink]{cleveref}
\Crefname{paragraph}{Section}{Sections}
\definecolor{egtgreen}{RGB}{75, 155, 8}
\definecolor{egtpurple}{RGB}{10, 33, 128}
\definecolor{egtred}{RGB}{103, 25, 15}
\definecolor{burgundy}{rgb}{0.5, 0.0, 0.13}
\definecolor{cyanp}{RGB}{45,129,173}
\hypersetup{colorlinks=true,linkcolor={egtred},citecolor=green!50!black,bookmarksnumbered=true}
\hypersetup{%
  pdfcreator={},
  pdfsubject={},%
  pdfproducer={},%
pdfinfo={%
  Title={Capturing Misalignment},%
  Author={Pf. Guarino \& Gabriel Ziegler},%
  ModDate={ },%
  CreationDate={ }%
}%
}
\newcommand{\Mref}[2][cyanp]{%
\hypersetup{linkcolor=#1}%
\Cref{#2}%
\hypersetup{linkcolor=burgundy}%
}
\newcommand{\Sref}[2][burgundy]{%
\hypersetup{linkcolor=#1}%
\Cref{#2}%
\hypersetup{linkcolor=burgundy}%
}
\usepackage{xpatch}\xpatchbibmacro{pageref}{parens}{brackets}{}{}
\makeatletter
\patchcmd{\BR@backref}{\newblock}{\newblock[}{}{}
\patchcmd{\BR@backref}{\par}{]\par}{}{}
\makeatother
\theoremstyle{plain}
\newtheorem{theorem}{Theorem}
\newtheorem{definition}{Definition}[section]
\newtheorem{proposition}[theorem]{Proposition}
\newtheorem{lemma}{Lemma}[section]

\newtheorem{remark}{Remark}[section]
\theoremstyle{definition}
\theoremstyle{remark}
\makeatletter
\def\thm@space@setup{%
  \thm@preskip=0.5cm 
  \thm@postskip=\thm@preskip 
}
\makeatother
\newtheoremstyle{myclaimstyle} 
    {\topsep}                    
    {\topsep}                    
    {\itshape}                   
    {}                           
    {\itshape}                   
    {.}                          
    {.5em}                       
    {}  

\theoremstyle{myclaimstyle}

\newtheorem{claim}{Claim}
\newtheorem*{myclaim}{Claim}

\makeatletter
\preto\claim{%
  \patchcmd\cref@thmnoarg
    {\trivlist}
    {\list{}{\leftmargin\parindent\rightmargin\parindent}}
    {}{}%
  \patchcmd\cref@thmoptarg
    {\trivlist}
    {\list{}{\leftmargin\parindent\rightmargin\parindent}}
    {}{}%
  \patchcmd\thmt@original@endclaim{\endtrivlist}{\endlist}{}{}%
}
\makeatother

\makeatletter
\preto\myclaim{%
  \patchcmd\cref@thmnoarg
    {\trivlist}
    {\list{}{\leftmargin\parindent\rightmargin\parindent}}
    {}{}%
  \patchcmd\cref@thmoptarg
    {\trivlist}
    {\list{}{\leftmargin\parindent\rightmargin\parindent}}
    {}{}%
  \patchcmd\thmt@original@endmyclaim{\endtrivlist}{\endlist}{}{}%
}
\makeatother
\newtheoremstyle{named}{}{}{\itshape}{}{\bfseries}{.}{.5em}{\thmnote{#3}}
\theoremstyle{named}

\makeatletter
\newtheoremstyle{axioms}{}{}{\itshape}{}{\bfseries}{.}{.5em}%
{#1 \thmnumber{#2} \@ifnotempty{#3}{ (\thmnote{#3})}}
\makeatletter
\newtheoremstyle{case}
  {5pt}
  {5pt}
  {\addtolength{\@totalleftmargin}{0em}
   \addtolength{\linewidth}{-1em}
   \parshape 1 1em \linewidth}
  {}
  {\normalfont}
  {.}
  {.5em}
  {}
\makeatother

\theoremstyle{case}

\AtEndEnvironment{proof}{\setcounter{step}{0}}
\AtEndEnvironment{proof}{\setcounter{claim}{0}}
\AtEndEnvironment{proof}{\setcounter{Cases}{0}}
\makeatletter
\def\thm@space@setup{%
  \thm@preskip=0.5cm 
  \thm@postskip=\thm@preskip 
}
\makeatother
\usepackage{thmtools}

\declaretheorem[numbered=yes, style=definition, name=Example, qed=$\diamond$]{example}
\renewcommand\thmcontinues[1]{Continued}
\declaretheoremstyle[
  spaceabove=3pt, spacebelow=6pt,
  headfont=\normalfont\itshape,
  notefont=\mdseries,
  bodyfont=\normalfont,
  postheadspace=.5em
]{innerproof}

\renewcommand\thmcontinues[1]{}

\declaretheorem[style= innerproof, numbered=unless unique, name=Proof of Claim, qed=$\square$]{proofclaim}
\renewcommand\thmcontinues[1]{}
\makeatletter
\preto\proofclaim{%
  \patchcmd\cref@thmnoarg
    {\trivlist}
    {\list{}{\leftmargin\parindent\rightmargin\parindent}}
    {}{}%
  \patchcmd\cref@thmoptarg
    {\trivlist}
    {\list{}{\leftmargin\parindent\rightmargin\parindent}}
    {}{}%
  \patchcmd\thmt@original@endproofclaim{\endtrivlist}{\endlist}{}{}%
}
\makeatother

\usepackage{datetime}
\makeatletter
\newcommand\Romanmonth{\@Roman{\month}}
\makeatother
\newdateformat{mydate}{\THEDAY.\Romanmonth.\THEYEAR}
\title{\vspace{-1.3cm}Capturing Misalignment%
\thanks{We would like to thank the audiences of the 2024 Spring Workshop on Economic Theory (at Bocconi), of the \emph{Séminaire parisien de Théorie des Jeux} at the Institut Henri Poincaré (Paris), of the SAET2025 Conference (Ischia), of the Soleto 2025 Workshop \Say{Experimetrics \& Behavioral Economics}, and of the seminars held at the University of Rome \Say{Tor Vergata},  the University of Siena, and the École Normale Supérieure Paris-Saclay. Of course, all errors are ours. Pierfrancesco thankfully acknowledges partial support from the Departmental Strategic Plan (PSD -- 2022-2025) of the Department of Economics and Statistics (DIES) of the University of Udine and hospitality at the Centre for Economics at Paris-Saclay (CEPS) of the École Normale Supérieure Paris-Saclay. Gabriel thankfully acknowledges financial support through the BA/Leverhulme Small Research Grants (SRG2425) and hospitality at LUISS Guido Carli, Rome.}%
\\ 
}
\author{%
Pierfrancesco Guarino\thanks{%
University of Udine (Department of Economics and Statistics -- DIES). %
\textit{E-mails:} \texttt{pf.guarino@hotmail.com} \& %
\texttt{pierfrancesco.guarino@uniud.it}.}
{\ \& }%
Gabriel Ziegler\thanks{%
 	University of Edinburgh (School of Economics). %
 	\textit{E-mail:} \texttt{ziegler@ed.ac.uk}.}%
}
\date{\mydate\today}
\begin{document}


\renewcommand\thmcontinues[1]{Continued}

\maketitle


\vspace{-0.5cm}

\begin{abstract}
\noindent
We introduce and formalize misalignment, a phenomenon of interactive environments perceived from an analyst’s perspective where an agent holds beliefs about another agent’s beliefs that do not correspond to the actual beliefs of the latter. We demonstrate that standard frameworks, such as type structures, fail to capture misalignment, necessitating new tools to analyze this phenomenon. To this end, we characterize misalignment through non-belief-closed state spaces and introduce agent-dependent type structures, which provide a flexible tool to understand the varying degrees of misalignment. Furthermore, we establish that appropriately adapted modal operators on agent-dependent type structures behave consistently with standard properties, enabling us to explore the implications of misalignment for interactive reasoning. Finally, we show how speculative trade can arise under misalignment, even when imposing the corresponding assumptions that rule out such trades in standard environments.

\bigskip


\end{abstract}

\section{Introduction}
\label{sec:introduction}

\subsection{Motivation \& Results}
\label{subsec:motivation_results}

Misalignment is a pervasive---yet unexplored---phenomenon of interactive environments, where it is typically assumed that agents hold beliefs not only about uncertain outcomes but also about the beliefs of others. Misalignment arises when the beliefs of an agent about another agent’s beliefs fail to correspond to the actual beliefs of the latter, as considered (or actually elicited) by an external analyst. 

Although the informal definition of misalignment provided above may appear convoluted, the underlying idea is not only highly intuitive, but also ubiquitous. To illustrate this point, we now present---informally at this stage---a scenario that serves as our recurring example throughout this paper.

\begin{example}[label=ex:misalignment, name=Weather in Paris \& Misalignment]
Consider two agents, namely, Ann and Bob, and a common domain of uncertainty which captures the idea that tomorrow it rains or it does not in Paris. Now, we assume that, as analysts, the following is the---very basic---scenario we contemplate: 
\begin{enumerate}[label=$\bullet$, leftmargin=*]
\item Ann assigns probability $1$ to the event that it will rain in Paris tomorrow, and to the belief that Bob also assigns probability $1$ to the same event, and further to the belief that Ann herself believes this, continuing \emph{ad infinitum}, where the beliefs just described are the \emph{only} ones Ann holds;

\item however (and rather crucially), this is simply not true, because the \emph{only} belief Bob holds is that he assigns probability $1$ to the event that it will not rain in Paris tomorrow, and that Ann shares his opinion \emph{and} assigns probability $1$ to the belief that Bob holds this view, continuing \emph{ad infinitum}.
\end{enumerate}
Hence, in what we just sketched, misalignment arises essentially at the second level of the belief hierarchies for both agents. For example, focusing on Ann, we know that her second-order belief puts probability $1$ on the event that it will rain in Paris tomorrow \emph{and} that Bob puts probability $1$ on the event that it will rain in Paris. However, we know Bob's first-order belief puts probability $1$ on the event that it will not rain in Paris tomorrow, contrary to the belief Ann attributes him. For clarification purposes, it should be pointed out that what we have in this example is an \say{extreme} form of misalignment, where, focusing on Ann, she puts probability $1$ on the \say{wrong} first-order belief of Bob. However, to have misalignment in the context of this example, it would have been enough for Ann to simply put positive probability (not necessarily equal to $1$) to the belief that Bob's first order belief puts probability $1$ on the event that it will rain in Paris (i.e., on what is \emph{not} an actual first-order belief of Bob).
\end{example}

If we want to capture this phenomenon along with its corresponding behavioral predictions, care has to be taken. Indeed, such mismatches challenge the standard frameworks used to model interactive reasoning, which implicitly assume alignment across agents’ hierarchies of beliefs: contemplating misalignment leads to explicitly introduce subtle discrepancies between different agents’ hierarchies of beliefs that elude conventional modeling tools like type structures. Addressing this complexity requires developing a novel theoretical framework capable of capturing the intricate structure of higher-order beliefs and their interactions under misalignment.

At its core, misalignment exhibits three defining properties:
\begin{enumerate}[leftmargin=*,label=\arabic*)]
\item it arises only in \emph{interactive environments}, where multiple agents are involved;

\item it emerges across agents’ \emph{hierarchies of beliefs}, reflecting higher-order reasoning about each other;

\item it is \emph{elicitable} in principle, meaning its presence can be identified and analyzed through appropriate tools.
\end{enumerate}
Thus, in this paper, we provide a formal definition of misalignment true to the properties identified above, by additionally providing a formal foundation for understanding this phenomenon, demonstrating its implications for economic theory, and offering tools to systematically model and analyze its effects, to then subsequently demonstrate that incorporating it into economic analyses can yield novel behavioral predictions. As such, we show that misalignment is not simply interesting \emph{per se} from a theoretical standpoint without a \say{practical} counterpart; rather, it is a significant phenomenon that must be incorporated into economics more broadly, since accounting for its presence enhances our behavioral predictions.

For the purpose of formalizing misalignment, by appropriately taking into account Points (1)--(3) above, we opt for infinite coherent hierarchies of beliefs---as formalized in \cite{Mertens_Zamir_1985} among others---as the natural tool for the task at hand. Now, types in the sense of \cite{Harsanyi_1967-1968} belonging to a type structure and a corresponding state space not only are the standard tool employed in economics to implicitly capture infinite coherent hierarchies of beliefs, but they \emph{are} infinite coherent hierarchies of beliefs as shown in the aforementioned \cite{Mertens_Zamir_1985}. Thus, building on this, we define misalignment as a property of state spaces (as defined in \Mref{def:state_space}) and for this we recall in \Sref{sec:background_knowledge} the definition of type structures and the construction of the Canonical Hierarchical Type Structure in the spirit of \cite{Mertens_Zamir_1985}: this structure, alternatively called the \Say{Universal\footnote{\label{foot:universal}The term \Say{universal} is actually employed in the literature in a technical sense: see \citet[Sections 3.4 \& 5.2.3]{Guarino_2025} on this point. Also, starting with topological assumptions as we do in this work, the uniqueness of this object is up to homeomorphism.} Type Structure}, is the \emph{unique} type structure that contains \emph{all} the possible interactive beliefs, given a set of agents and a domain of uncertainty.

Reviewing these topics is crucial, because it demonstrates why \say{standard} type structures \emph{à la} \cite{Harsanyi_1967-1968} \emph{fail} to capture misalignment and why it is necessary to introduce an appropriate framework to address its presence. In particular, the problem that arises when employing \say{standard} type structures can be described via \Mref{ex:misalignment} by observing that, if we aim to capture the infinite coherent hierarchies of beliefs held by Ann and Bob that we, as analysts, deem relevant, we need:
\begin{itemize}[leftmargin=*]
\item a type for Ann, denoted by $\type^{\rain}_\Ann$, \emph{and} a type for Bob, denoted by $\type^{\rain}_\Bob$, where $\type^{\rain}_\Ann$ assigns probability $1$ to the event that it will rain, and to Bob being of type $\type^{\rain}_\Bob$, \emph{and} $\type^{\rain}_\Bob$ assigns probability $1$ to the event that it will rain, and to Ann being of type $\type^{\rain}_\Ann$;

\item a type for Bob, denoted by $\type^{\sun}_\Bob$, \emph{and} a type for Ann, denoted by $\type^{\sun}_\Ann$, where $\type^{\sun}_\Bob$ assigns probability $1$ to the event that it will not rain, and to Ann being of type $\type^{\sun}_\Ann$, \emph{and} $\type^{\sun}_\Ann$ assigns probability $1$ to the event that it will rain, and to Bob being  type $\type^{\sun}_\Bob$.
\end{itemize}

Building on these two points, we can construct a \say{standard} type structure and a corresponding state space based on the types described above, which actually represent nothing else but infinite coherent hierarchies of beliefs. This would lead to the following crucial problem, though: for example, when focusing on Ann, this framework fails to distinguish between $\type^{\rain}_\Ann$ and $\type^{\sun}_\Ann$, since these two types have the same \say{weight} in the analysis that can be performed within it. However, these types \emph{are} actually different: as analysts, we \emph{know} that $\type^{\rain}_\Ann$ represents the \emph{only} infinite coherent hierarchy of beliefs held by Ann, whereas $\type^{\sun}_\Ann$ is merely a by-product of the framework required to capture, via types, the \emph{only} infinite coherent hierarchy of beliefs held by Bob, which is captured by $\type^{\sun}_\Bob$.\footnote{Of course, the same point applies \emph{mutatis mutandis} to $\type^{\rain}_\Bob$ for Bob, which is a by-product of the framework required to capture the \emph{only} infinite coherent hierarchy of beliefs held by Ann, captured by $\type^{\rain}_\Ann$.}

In this example, if we as analysts are interested in accounting for the potential presence of misalignment we should \emph{not} consider the state space of interest to be the one that contains $\type^{\rain}_\Ann$, $\type^{\sun}_\Ann$, $\type^{\rain}_\Bob$, and $\type^{\sun}_\Bob$. Rather, care must be taken in determining the state space of interest and in this example we should focus on a state space that contains only $\type^{\rain}_\Ann$ and $\type^{\sun}_\Bob$, since---after all---these represent the \emph{only} infinite coherent hierarchies of beliefs that we, as analysts, consider relevant. Thus, building on this intuition and using the tools described in \Sref{sec:background_knowledge}, we provide a formal definition of misalignment in \Mref{def:misaligned_state_space} as a property of a state space consistent with the informal definition provided in the opening paragraph above.

Now, not surprisingly, the resulting definition is as intricate as the informal one. Thus, this necessitates a characterization that simplifies the notion and makes it more tractable. Therefore, we provide a fully tractable characterization that makes misalignment easy to handle in the context of state spaces, as detailed in \Mref{th:state_space_characterization}, which is a result of independent technical interest. Indeed, \Mref{th:state_space_characterization} shows that a state space is misaligned if and only if it is non-belief-closed (where the notion of belief-closed state space is provided in \Mref{def:belief-closed_state_space}). In addition to greatly simplifying the verification of whether a state space is misaligned, this result might be of independent interest, as it marks the first time that non-belief-closed state spaces take center stage in the literature. To the extent that, had this work focused differently on the technical aspects of interactive epistemology, this result may alternatively be regarded as a standalone characterization of non-belief-closed state spaces.\footnote{Although it aligns closely with some arguments made in \citet{Mertens_Zamir_1985}.}

Whereas \Mref{th:state_space_characterization} is conceptually appealing at large, it leads to another problem: it is inherently infeasible to perform meaningful analysis with non-belief-closed state spaces due to the inherent limitations of the tools used in economics and game theory more broadly, and interactive epistemology in particular. To resolve this issue, we must transform a misaligned (and non-belief-closed) state space into one that is belief-closed. However, this cannot be achieved by simply collecting all the types that are \say{missing} and then adding them to the misaligned state space, as this would encounter the problem outlined above, where \emph{all} types are assigned the same \say{weight} for analytical purposes. Consequently, we introduce in \Mref{def:agent-dependent_type_structure} the notion of an agent-dependent type structure. More specifically, an agent-dependent type structure is a \say{standard} type structure that is \emph{a fortiori} belief-closed (in light of \Mref{rem:belief-closed_type_structure}) and contains the infinite coherent hierarchies of beliefs that we, as analysts, consider relevant. In essence, an agent-dependent type structure formalizes the infinite coherent hierarchies of beliefs that reside in the mind of a given agent. Armed with this notion, and given a set of agents and a state space of interest, we can construct a \emph{profile} of agent-dependent type structures, one for each agent, which enables us to conduct a more nuanced analysis. Indeed, in \Mref{def:degenerate_agent-dependent_type_structure} we deem a profile of agent-dependent type structures:
\begin{enumerate}[leftmargin=*,label=$\bullet$]
\item \emph{degenerate} if no agent-dependent type structure contains the types of an agent---as infinite coherent hierarchies of beliefs---beyond those initially given (otherwise, it is non-degenerate);

\item \emph{common} if, for any pairwise comparison of the state spaces derived from two agent-dependent type structures in the profile, the state spaces are equal (otherwise, it is non-common).
\end{enumerate}
Thus, these definitions lead to a taxonomy (represented by a $2 \times 2$ matrix, as shown in \Sref{tab:misalignment_taxonomy}) which reveals hidden assumptions embedded in state spaces. In particular, it turns out that the \say{standard} type structures (along with the corresponding state spaces) employed in the literature are a very special case of a profile of agent-dependent type structures, namely, the specific case in which the profile of agent-dependent type structures is both \emph{common} and \emph{degenerate}.

Additionally, we introduce the notion of a \emph{minimal} agent-dependent type structure, as defined in \Mref{def:minimal_agent-dependent_type_structure}. This is the agent-dependent type structure which contains \emph{only} those types strictly required to construct a belief-closed state space from the infinite coherent hierarchies of beliefs of interest for a given agent (as captured via types). Additionally, we show that, starting from a set of agents and a state space of interest, we can always construct a space called the \Say{Agent Closure} (as defined in \Mref{def:agent_closure}), which contains only those types that originate from the infinite coherent hierarchies of beliefs of interest for a given agent (as captured via types). We further show in \Mref{th:characterization_minimality} that the agent-dependent type structure derived from the agent closure is well-defined (i.e., it exists, is unique, and belief-closed) and is indeed the minimal one.

Now, after resolving the conceptual, technical, and foundational issues arising when trying to capture misalignment, we are now in a position to undertake interactive epistemological analysis under misalignment. For this purpose, we adapt standard modal operators to account for the presence of misalignment, and we show in \Mref{prop:real_CB_measurability} and \Mref{eq:misalignment_belief_operator_properties} that these operators satisfy the usual properties. Specifically, the application of these modal operators relies on employing the \say{standard} modal operators on the agent-dependent state spaces, and then \emph{intersecting} the results with the infinite coherent hierarchies of beliefs of interest.

Finally, this positions us to address the issue of behavioral predictions arising from the presence of misalignment, as discussed in \Sref{sec:misalignment_matters}. For this purpose, we focus in this work on a class of results that serve as both a benchmark and a test for works dealing with state spaces, i.e., No-Speculative Trade theorems \emph{à la} \citet[Theorem 1, p.21]{Milgrom_Stokey_1982}. It turns out that particular care must be taken with the notion of a \emph{common prior}, which plays a crucial role in these results,\footnote{See \Sref{subsec:conceptual_aspects} \Say{Common Priors \& Heterogeneous Priors} with respect to this point.} when we contemplate the presence of misalignment. Once this point is appropriately addressed, we show that the taxonomy introduced above, based on a profile of agent-dependent type structures, proves to be insightful. Indeed, on one hand, we obtain results analogous to \citet[Theorem 1, p.21]{Milgrom_Stokey_1982} in \Mref{th:generalized_no-speculative-trade_derived}. On the other hand, and crucially, we derive novel behavioral predictions from a profile of agent-dependent type structures that is both non-degenerate and non-common: specifically, in this case, even when assumptions of \citet[Theorem 1, p.21]{Milgrom_Stokey_1982} translated to our framework are satisfied, speculative trade can still occur.

One question arises naturally and can now be clearly stated and addressed: given how the introduced modal operators work, i.e., by intersecting the results obtained via \say{standard} modal operators with the infinite coherent hierarchies of beliefs of interest for the analyst, would we not obtain the same novel behavioral predictions mentioned above \emph{without} introducing agent-dependent type structures in the first place?\footnote{It should be observed that this question does not have any impact on the importance of misalignment \emph{per se}, but it simply concerns the path chosen in this work to obtain the behavioral predictions corresponding to having misalignment.} The answer is actually yes in the considered framework, but with a major drawback: we would not be able to understand \emph{why} those behavioral predictions arise. That is, it is the introduction of agent-dependent type structures, which---crucially---leads to the taxonomy described above, that allows us to pinpoint the reasons behind certain behavioral predictions. In other words, without the aforementioned taxonomy, we would not understand the actual driving force behind the behavioral predictions obtained under misalignment.

\subsection{Related Literature}
\label{subsec:related_literature}

Beyond the references alluded to previously, this work is related to the literature that aims to address various cognitive\footnote{\label{foot:cognitive}We use the term \Say{cognitive} in this context in a broad sense as everything which stems from cognition, where the latter is defined in the Encyclopedia Britannica as \emph{\Say{the states and processes involved in knowing, which in their completeness include perception and judgment. Cognition includes all conscious and unconscious processes by which knowledge is accumulated, such as perceiving, recognizing, conceiving, and reasoning}.}} phenomena, such as information processing mistakes, unawareness, or bounded reasoning, as in \cite{Geanakoplos_2021},\footnote{Working paper version dating back to 1989.} \cite{Heifetz_et_al_2006}, and \cite{Alaoui_Penta_2016}, with particular emphasis on mistakes regarding the beliefs of other agents, as explored in \cite{Bursztyn_Yang_2022}. In terms of the tools employed, this paper is related to \cite{Harsanyi_1967-1968}, \cite{Aumann_1976}, and \cite{Mertens_Zamir_1985}, which address the problem of capturing infinite coherent hierarchies of beliefs. However, this work represents a significant shift from these earlier contributions, as it is the first to explicitly address the problem of creating \say{personalized} type structures to capture all the infinite coherent hierarchies of beliefs that can arise in the \say{mind} of an agent. Specifically, this paper provides a formalization and foundation for the notion of player-specific type structures discussed in \citet[Section 8.1]{Brandenburger_Friedenberg_2010}.\footnote{Although used for very different questions, the framework put forward by \citet{Piermont_Zuazo_2024} can be seen as a different formalization of player-specific type structures.} Also, this work relates to the methodological point in \citet[Chapter 12.1.1, pp.622–624]{Dekel_Siniscalchi_2015}, which emphasizes the importance of addressing elicitability issues when dealing with types and infinite coherent hierarchies of beliefs. Finally, our approach sheds light on the usage of type structures in experiments based on games as in \cite{Kneeland_2015}.\footnote{See \Sref{subsec:conceptual_aspects} \Say{Misalignment \& Experiments on Games}.}

\subsection{Synopsis}
\label{subsec:synopsis}

The remainder of the paper is organized as follows. In \Sref{sec:background_knowledge}, we provide a self-contained introduction to \emph{type structures} and demonstrate how they can be used to model interactive beliefs. In \Sref{sec:defining_misalignment}, we define and characterize \emph{misalignment}, while in \Sref{sec:a_framework_for_misalignment}, we introduce \emph{agent-dependent type structures} and provide their foundation. \Sref{sec:interactive_epistemology_under_misalignment} introduces appropriate modal operators to deal with the presence of misalignment and enable interactive epistemological analysis. Building on this, \Sref{sec:misalignment_matters} explores how misalignment can lead to novel behavioral predictions. Finally, in \Sref{sec:discussion}, we discuss various conceptual and technical aspects of our framework, along with potential directions for future research. All proofs of the results in the paper are relegated to \Sref{app:proofs}.

\section{Background Knowledge}
\label{sec:background_knowledge}

Throughout the paper, we fix a finite set $\Agents$ of \emph{agents}. Letting $0$ denote a dummy agent, referred to as \Say{Nature}, such that $0 \notin \Agents$, we define $\Agents_0 := \Agents \cup \{0\}$. From a topological standpoint, all spaces are assumed to be compact and implicitly endowed with their Borel $\sigma$-algebra. Given an arbitrary (compact) product space $\ArbSet := \prod_{j \in \Agents} \ArbSet_j$, adopting standard game theoretical conventions, we define $\ArbSet_{-i} := \prod_{j \in \Agents \setminus \{i\}} \ArbSet_j$, with similar notation used for elements of product spaces. In either case, $\proj$ denotes the projection operator as canonically defined. Product spaces are endowed with the product topology and the associated product Borel $\sigma$-algebra. We denote by $\Delta (\ArbSet)$ the space of all $\sigma$-additive probability measures (henceforth, probability measures) over an arbitrary compact space $\ArbSet$. In this case, $\Delta (\ArbSet)$ is itself endowed with the topology of weak convergence, making $\Delta(X)$ compact. For product spaces, we denote the marginal operator by $\marg$, as canonically defined. Finally, we let $\Naturals := \Sets { 1, 2, \dots}$ with $\Naturals_0 := \Naturals \cup \{0\}$.

Starting from a (compact) space $\SNatures$ which denotes a common domain of uncertainty of \emph{states of nature}, the set of \emph{infinite coherent hierarchy of beliefs} (henceforth, CHBs) is recursively defined for every $i \in \Agents$ as follows, with $m \in \Naturals$: 
\begin{center}
	\begin{tabular}{cc}
		$\States^{1}_i := \SNatures$, 		&%
		$\Hiers^{1}_i :=  \Delta (\States^{1}_{i})$, \\
		$\vdots$			& 	$\vdots$ \\
		$\States^{m+1}_i := \SNatures \times \Hiers^{m}_{-i}$, &%
		$\Hiers^{m+1}_i := \Set { ((\mu^1_i, \ldots, \mu^{m}_i),\mu^{m+1}_i) \in \Hiers^{m}_i \times \Delta \big(\States^{m+1}_i\big) | \marg_{\States^{m}_i} \mu^{m+1}_i = \mu^{m}_i }$,   \\
		$\vdots$		&	 	$\vdots$ \\
	\end{tabular}
\end{center}
with $\Hiers^{\ell} := \prod_{j \in \Agents} \Hiers^\ell_j$ for every $\ell \in \Naturals$. Considering $i \in I$ and letting $\invlim (\Hiers^{\ell}_i)_{\ell \in \Naturals}$ denote the limit of the inverse system\footnote{See \citet[Chapter 2.5, p.98]{Engelking_1989}} of spaces  $(\Hiers^{\ell}_i)_{\ell \in \Naturals}$, the compact space
\begin{equation*}
\UTypes_i := \invlim (\Hiers^{\ell}_i)_{\ell \in \Naturals}
\end{equation*}
denotes the \emph{universal type space of agent $i$}, where a type of an agent is a CHB that admits an extension to the limit.\footnote{Starting from topological assumptions (as we do here) with the corresponding Borel $\sigma$-algebras, agents' types \emph{are} CHBs as a consequence of Kolmogorov's Extension Theorems like results. However, in the purely measurable (topology-free) case, the universal type space of an agent is a \emph{proper} subset of the space of that agent's CHBs, as shown in \citet[Section 4]{Heifetz_Samet_1999} (see \cite{Fukuda_2024} for a construction that delivers the usual equivalence).} Then $\UTypes := \prod_{j \in \Agents} \UTypes_j$ denotes the \emph{universal type space} and 
\begin{equation}
\label{eq:universal_state_space}
\UStates := %
\SNatures \times \UTypes = \SNatures \times \prod_{j \in \Agents} \UTypes_j
\end{equation}
is the (compact) \emph{universal state space}. To go from CHBs to specific levels of the hierachy, we implicitly define, for a given agent $i \in \Agents$ and an $m \in \Naturals$, the following functions:\footnote{See \citet[Section 2, pp.4--6]{Mertens_Zamir_1985} for details.} the function
\begin{align}
\label{eq:projection_state}
\begin{alignedat}{2}
    \projstfun^{m}_i : \, \UTypes_i \, & \longrightarrow \, \Hiers^{m}_i ,\\
    \type_i^* = (\mu_i^1, \ldots, \mu_i^m, \ldots) \,
        &\longmapsto \, \type_i^m=(\mu_i^1, \ldots, \mu_i^m)
\end{alignedat}
\end{align}
denotes the continuous \emph{$m^{\text{th}}$-order type projection function}, whereas the function
\begin{align}
\label{eq:hierarchy_function}
\begin{alignedat}{2}
    \hierfun^{m}_i  : \,\UTypes_i \, &\longrightarrow \,  \Delta  (\States^{m}_{i})=\Delta  (\SNatures \times \Hiers^{m-1}_{-i}) ,\\
    \type_i^* = (\mu_i^1, \ldots, \mu_i^m, \ldots) \,
        & \longmapsto \, \mu_i^m
\end{alignedat}
\end{align}
denotes the continuous \emph{$m^{\text{th}}$-level universal hierarchy function}.

Importantly, \citet[Theorem 2.9, pp.7--8]{Mertens_Zamir_1985} establish the existence of a unique homeomorphism\footnote{See also \citet[Theorem 4.2, p.19]{Armbruster_Boge_1979}, \citet[Theorem 9, p.333]{Heifetz_1993}, and \citet[Proposition 2, p.193]{Brandenburger_Dekel_1993}, where the latter is the result of a different construction with coherency imposed \emph{after} the construction of infinite hierarchies of beliefs not necessarily coherent.} 
\begin{equation}
\label{eq:type_homeomorphism}
\belhom_i : \UTypes_i \to \Delta (\SNatures \times \UTypes_{-i} )
\end{equation}
for every $i \in \Agents$. For our purposes here, we introduce the following derived continuous function
\begin{align*}
\begin{alignedat}{2}
    \typehom_i  : \, \UTypes_i \, & \longrightarrow \,  \Delta (\UTypes_{-i}) ,\\
    \type_i^* & \longmapsto \, \marg_{\UTypes_{-i}} \belhom_i (\type^*_i)
\end{alignedat}
\end{align*}
for every $i \in \Agents$. 

Collecting the key objects introduced above gives rise to the \emph{Canonical Hierarchical Type Structure} (alternatively, the Universal\footnote{See \Sref{foot:universal} concerning this terminology.} Type Structure) on $\SNatures$, defined as the tuple:
\begin{equation}
\label{eq:canonical_hierarchical_type_structure}
\THierSt := \Angles { \Agents , \SNatures , \Rounds { \UTypes_j , \belhom_j  }_{j \in \Agents} } .
\end{equation}
By construction, the Canonical Hierarchical Type Structure contains \emph{all} CHBs for a given set of agents and a domain of uncertainty $\SNatures$. However, for modeling purposes, it is not always necessary to employ such a rich object. In particular, it is often possible to focus on \say{small} counterparts of the Canonical Hierarchical Type Structure that restrict the CHBs considered in the analysis. 

Thus, a \emph{(product) type structure} on $\SNatures$ (henceforth referred to as a \say{standard} type structure)\footnote{In the literature, these objects are simply called \Say{type structures}. We add the adjective \Say{standard} to distinguish them from the objects we introduce in \Sref{sec:a_framework_for_misalignment}.} is a tuple

\begin{equation}
\label{eq:type_structure}
\TypeSt := %
\Angles { \Agents, \SNatures , \Rounds {\Types_j, \belfun_j}_{j \in \Agents} }
\end{equation}
where, for every $i \in I$, agent $i$'s \emph{type space} $\Types_i$ is a compact set and $\belfun_i : \Types_i \to \Delta (\SNatures \times \Types_{-i})$ is a continuous \emph{belief function}.

Given a \say{standard} type structure $\TypeSt$, by usual arguments it is possible to retrieve the CHB corresponding to any type $\type_i \in \Types_i$ for every $i \in \Agents$ using the \emph{hierarchy function} of agent $i$. If all these hierarchy functions are injective, these functions embed every \say{standard} type structure into the Canonical Hierarchical Type Structure in a belief-preserving manner. Since, in this work, we are concerned only with CHBs and not with types \emph{per se}, we implicitly assume that all \say{standard} type structures are non-redundant and---by abusing notation \say{forgetting} the hierarchy function---we consider $\Types_i \subseteq \UTypes_i$ henceforth, which naturally gives rise to a partial order on \say{standard} type structures. The following remark summarizes this discussion and therefore, for the remainder of this paper, every type in a \say{standard} type structure \emph{is} a CHB.

\begin{remark}[Non-Redundancy, Order \& Notation]
\label{rem:subset_of_universal}
Given a \say{standard} type structure 
\begin{equation*}
\TypeSt := %
\Angles { \Agents, \SNatures , \Rounds { \belfun_j , \Types_j}_{j \in \Agents} } ,
\end{equation*}
$\TypeSt$ is assumed to be non-redundant with $\Types_i \subseteq \UTypes_i$ and $\belfun_i={\belhom_i}\vert_{ T_i}$, for every $i \in \Agents$. Given another \say{standard} type structure
\begin{equation*}
\TypeSt^\prime := %
\Angles { \Agents, \SNatures , \Rounds { \belfun^\prime_j , \Types^\prime_j}_{j \in \Agents} } ,
\end{equation*}
write $\TypeSt \subseteq \TypeSt^\prime$ if $\Types_i \subseteq \Types_i^\prime$, for every $i \in I$.
\end{remark}

Now, in this paper, contrary to the standard definition that can be found in the literature, we define a \emph{state space} and a \emph{belief-closed state space} as follows.

\begin{definition}[State Space]
\label{def:state_space}
A \emph{state space} $\States := \SNatures \times \prod_{j \in \Agents} \Types_j$ is a nonempty compact (hence, measurable) subspace of the universal state space $\UStates = \SNatures \times \UTypes$. Moreover, given a \say{standard} type structure 
\begin{equation*}
\TypeSt := %
\Angles { \Agents, \SNatures , \Rounds { \belfun_j , \Types_j}_{j \in \Agents} } ,
\end{equation*}
$\TypeSt$ \emph{induces a state space} $\States$ by setting $\States := \SNatures \times \prod_{j \in \Agents} \Types_j$.
\end{definition}

\begin{definition}[Belief-Closed State Space]
\label{def:belief-closed_state_space}
A state space $\States := \SNatures \times \prod_{j \in \Agents}\Types_j$  is \emph{belief-closed} if 
\begin{equation}
\label{eq:belief-closedness}
\supp \typehom_i (\type_i) \subseteq \Types_{-i}
\end{equation}
for every $i \in \Agents$ and $\type_i \in \Types_i$. 
\end{definition}

Two points should be emphasized regarding these two definitions, which are closely related. First, our notion of state space, as defined in \Mref{def:state_space}, is new and, unlike the existing literature, it does \emph{not} require the state space to be belief-closed. Second, the notion of \Mref{def:belief-closed_state_space} that we employ is new and serves as an adapted translation of the original definition in \citet[Definition 2.15, p.12]{Mertens_Zamir_1985} to the context of \say{standard} type structures:\footnote{\label{foot:belief_structures}\cite{Mertens_Zamir_1985} work with \emph{belief structures} (in their terminology, belief spaces). These structures generalize \say{standard} type structures because they lack a product structure initially but are homeomorphic to (product) \say{standard} type structures. See also \citet[Chapters 10 \& 11]{Maschler_et_al_2013}.} specifically, in \say{standard} type structures, what determines whether a state space is belief-closed are simply the beliefs of every agent's type regarding the types of the other agents \emph{alone}.\footnote{Thus, in the context of \say{standard} type structures, we could, in principle, use the expression \Say{belief-closed type space}.}

Furthermore, keeping in mind \Mref{rem:subset_of_universal}, a belief-closed state space $\States = \SNatures \times \prod_{j \in \Agents} \Types_j$ induces a corresponding \say{standard} type structure by defining $\belfun_i (\type_i) := \belhom_i (\type_i)$ for every $i \in \Agents$ and $\type_i \in \Types_i$. These observations are summarized in the following remark:

\begin{remark}
\label{rem:belief-closed_type_structure}
A \say{standard} type structure induces a belief-closed state space and vice versa.
\end{remark}

We now demonstrate how the framework introduced can be applied to capture interactive reasoning in the scenario described in \Sref{sec:introduction}.

\begin{example}[continues=ex:misalignment, name=Interactive reasoning via \say{Standard} Type Structures]
We start by defining $\Agents := \Sets { \Ann , \Bob }$, with \Say{$\Ann$} standing for Ann and \Say{$\Bob$} standing for Bob, and $\SNatures := \Sets { \snature^{\rain} , \snature^{\sun} }$, with \Say{$\snature^{\rain}$} standing for having rain in Paris and \Say{$\snature^{\sun}$} standing for not having rain. Now, recalling the previously described story:
\begin{enumerate}[label=$\bullet$, leftmargin=*]
\item Ann assigns probability $1$ to it raining in Paris tomorrow and to Bob assigning probability $1$ to it raining \emph{and} to Ann assigning probability $1$ to it raining in Paris, continuing \emph{ad infinitum}. These beliefs are also shared by Bob;

\item Bob assigns probability $1$ to it not raining in Paris tomorrow and to Ann sharing his belief \emph{and} to her assigning probability $1$ to him assigning probability $1$ to it raining, continuing \emph{ad infinitum}. These beliefs are also shared by Ann.
\end{enumerate}
Letting $\Types_i := \{\type^{\rain}_i, \type^{\sun}_i\}$ for every $i \in \Agents$ with $\widetilde{\States} := \SNatures \times \Types_\Ann \times \Types_\Bob$, and we obtain a \say{standard} type structure which captures the described scenario by setting\footnote{We abuse notation by omitting set notation for singleton sets.} 
\begin{equation}
\label{eq:measure_convention}
\belfun_i (\type^{\weather}_i) := 1 (\snature^{\weather} , \type^{\weather}_{-i}) .
\end{equation}
for every $\weather \in \{\rain, \sun\}$ and $i \in \Agents$, where here and in what follows---modulo details---the RHS of \Mref{eq:measure_convention} stands for the probability measure $\belfun_i (\type^{\weather}_i) \in \Delta (\SNatures \times \Types_{-i})$ putting probability $1$ on the state-type pair $(\snature^{\weather} , \type^{\weather}_{-i})$. However, note that the scenario described above is actually formally represented \emph{only} by
\begin{align*}
\belfun_\Ann (\type^{\rain}_\Ann) & = 1 (\snature^{\rain} , \type^{\rain}_{\Bob}) \\
\belfun_\Bob (\type^{\sun}_\Bob) & = 1 (\snature^{\sun} , \type^{\sun}_{\Ann}) .
\end{align*}
Graphically, this interactive reasoning is illustrated in \Sref{fig:alignment_example}, where the players' CHBs that we---as omniscient outside observers---consider \say{real} are highlighted in red. Regarding the players' beliefs represented by arrows, a blue (resp., green) arrow moving from a state $\state$ to a state $\state'$ represents the fact that Ann's type at state $\state$ (resp., Bob's type) assigns probability $1$ to the pair $(\snature, \type_\Bob)$ (resp., $(\snature, \type_\Ann)$) belonging to $\state'$, with $\state, \state' \in \widetilde{\States}$, $\snature \in \SNatures$, $\type_\Ann \in \Types_\Ann$, and $\type_\Bob \in \Types_\Bob$.
\end{example}

\begin{figure}
\hspace{-2cm}
\begin{tikzpicture}
[real/.style={circle, draw, inner sep=1.5, fill=black},fake/.style={circle, draw, inner sep=1.5},scale=2] 
\node[real] (n_1) at (-0.5,0.5) {};
\node[real] (n_2) at (0.5,0.5) {};
\node[real] (n_3) at (-0.5,-0.5) {};
\node[real] (n_4) at (0.5,-0.5) {};
\draw (-0.7,0.7) node {$\state^1$}; 
\draw (0.7,0.7) node {$\state^2$};
\draw (-0.7,-0.7) node {$\state^{\rain}$};
\draw (0.7,-0.7) node {$\state^{\sun}$};
\draw[blue,->] (n_1) -- (n_3);
\draw[blue,->] (n_2) -- (n_3);
\path[blue,->] (n_3) edge  [loop left] node {} ();
\path[blue,->] (n_4) edge  [loop below] node {} ();
\draw[green,->] (n_1) -- (n_4);
\draw[green,->] (n_2) -- (n_4);
\path[green,->] (n_3) edge  [loop below] node {} ();
\path[green,->] (n_4) edge  [loop right] node {} ();
%
\matrix [below left] at (2.5,0.9) {
  \node [label=right:Ann] {}; \\
  \node [label=right:Bob] {}; \\
  \node [label=right:$\widetilde{\States}$] {}; \\
};
\draw[blue] (1.7,0.7) -- (2,0.7);
\draw[green] (1.7,0.48) -- (2,0.48);
\draw[] (1.7,0.2) -- (2,0.2);
\draw[rounded corners=7](-1,-1) rectangle (1,1);
\draw (-1.2,0.85) node {\Large$\widetilde{\States}$};
\draw (-3,0.7) node {$\state^1 := (\snature^\rain , \textcolor{red}{\type^{\rain}_\Ann} , \textcolor{red}{\type^{\sun}_\Bob} )$};
\draw (-3,0.4) node {$\state^2 := (\snature^\sun ,\textcolor{red}{\type^{\rain}_\Ann} , \textcolor{red}{\type^{\sun}_\Bob} )$};
\draw (-3,0.1) node {$\state^{\rain}  := (\snature^\rain , \textcolor{red}{\type^{\rain}_\Ann} , \type^{\rain}_\Bob )$};
\draw (-3,-0.2) node {$\state^{\sun} := (\snature^\sun , \type^{\sun}_\Ann , \textcolor{red}{\type^{\sun}_\Bob})$};
\end{tikzpicture}
\caption{A representation of the agents' beliefs in $\widetilde{\States}$.}
\label{fig:alignment_example}
\end{figure}

\section{Defining Misalignment}
\label{sec:defining_misalignment}

In \Sref{sec:introduction}, we informally defined misalignment along the following, now slightly more formal, lines: given at least two agents reasoning about a domain of uncertainty $\SNatures$ \emph{and} their respective beliefs about this domain (and so on \emph{ad infinitum}), misalignment exists if, for at least one agent $i$, there is an infinite hierarchy of beliefs in which, at some level, the agent holds a belief about another agent that does not match any belief actually held by that agent. 

Thus, the following definition provides an exact translation of what we just stated into the formalism introduced in \Sref{sec:background_knowledge}.

\begin{definition}[Misaligned State Space]
\label{def:misaligned_state_space}
A state space $\States =  \SNatures \times \prod_{j \in \Agents}\Types_j$ is \emph{misaligned} if there exists an agent $i \in \Agents$, a type $\type_i \in \Types_i$, an $m \in \Naturals \setminus \{1\}$, and an agent $j \in \Agents \setminus \{i\}$ such that 
\begin{equation}
\label{eq:misalignment}
 \Big( \supp \marg_{\Hiers^{m-1}_{j}} \hierfun^{m}_i (\type_i) \Big) \not\subseteq \projstfun^{m-1}_j \Types_j .
\end{equation}
\end{definition}

Although \Mref{def:misaligned_state_space} may appear cumbersome, it is merely a by-product of the formalism. First of all, we should observe that the left-hand side of \Mref{eq:misalignment} captures what agent $i$'s attributes to agent $j$, with the right-hand side capturing the beliefs (or higher-order beliefs) actually held by agent $j$. Thus, taking this point into account, \Mref{def:misaligned_state_space} expresses that a state space is misaligned if we can find an agent (i.e., agent $i$ in the definition) that at some level of one of their CHBs (captured via the function $\hierfun^{m}_i$ as defined in \Mref{eq:hierarchy_function}) holds beliefs about another agent's beliefs (obtained via the support and marginalization) that do not correspond to any beliefs that this other agent (i.e., agent $j$ in the definition) holds as they are captured in the relevant original state space (via the function $\projstfun^{m}_j$, as defined in \Mref{eq:projection_state}), i.e., they are not included in $\Types_j$.

The description above can alternatively be illustrated using the following diagram,\footnote{We are illustrating set inclusion via the corresponding surjection, denoted by the hooked arrow.} where the state space $\SNatures \times \Types_i \times \Types_j \times \Types_{-i,j} \subseteq \SNatures \times \UTypes_i \times \UTypes_j \times \UTypes_{-i,j}$ is misaligned if, starting from $\Types_i$, there exists a type $\type_i$ such that, by following the arrows clockwise to $\Types^{m-1}_j$ and taking the inverse of $\projstfun^{m-1}_j$, we end up with elements in $\UTypes_j$ but not in $\Types_j$:
\begin{equation}
\label{cd:misalignment}
\begin{tikzcd}[row sep = huge, column sep = huge]
{\SNatures \times \UTypes_i \times \UTypes_j \times \UTypes_{-i,j}} \arrow[rd, "{\proj^{m-1}_j}"', start anchor={[shift={(-0.5cm,0cm)}]}] & {\SNatures \times \Types_i \times \Types_j \times \Types_{-i,j}} \arrow[l, ""', hook'] \arrow[r, "{\hierfun^{m}_i}"] & {\Delta (\SNatures \times \Types^{m-1}_{-i})} \arrow[d, "{\marg_{\Types^{m-1}_j}}"] \\
                   & {\Types^{m-1}_j} \arrow[u, "{\big(\proj^{m-1}_j\big)^{-1}}"']                       & {\Delta(\Types^{m-1}_j)} \arrow[l, "\supp"]
\end{tikzcd}
\end{equation}
In particular, the diagram is commutative when there is no misalignment, whereas it is actually \emph{not} commutative in presence of misalignment, since it fails to satisfy commutativity on $\big(\proj^{m-1}_j\big)^{-1}$.

Hence, two aspects of \Mref{def:misaligned_state_space} are crucial. First, misalignment is a property of a (fixed) state space that we---as analysts---choose to employ as the \say{original} state space of interest for modeling interactive reasoning in a given scenario under scrutiny. Second, although misalignment is a property of a state space, it fundamentally concerns CHBs, as reflected in \Mref{eq:misalignment}, which is based on the notion of the $m^{\text{th}}$-order universal hierarchy function.

\begin{example}[continues=ex:misalignment, name=Uncovering the Definition]
Recall that we had $\Agents = \Sets { \Ann , \Bob }$ and $\SNatures := \Sets { \snature^{\rain} , \snature^{\sun} }$. Furthermore, since any type is a CHB in our notation, we also defined Ann's type $\type^{\rain}_\Ann$ to represent common certainty of rain, in accordance with the considered scenario. More formally, using the $m^{\text{th}}$-level universal hierarchy functions, we have:
\begin{align*}
\hierfun^{1}_\Ann (\type^{\rain}_\Ann) & =  1 (\snature^{\rain}) , \\
\hierfun^{m+1}_\Ann (\type^{\rain}_\Ann) & = %
1 \Big( 1 (\snature^{\rain}) , 1\big(\snature^{\rain} , 1(\snature^{\rain}) \big), %
\dots, %
\underbrace{1\big( 1 (\snature^{\rain}) , 1\big(\snature^{\rain} , 1(\snature^{\rain}) \big) \big), \dots}_{\text{$m+1$ times}}  \Big) ,
\end{align*}
for every $m > 1$. Similarly, Bob's type $\type^{\sun}_\Bob$ represents common certainty of \say{no rain}. Thus,
\begin{align*}
\label{eq:Bob_CHB}
\hierfun^{1}_\Bob (\type^{\sun}_\Bob) & =  1 (\snature^{\sun}) , \\
\hierfun^{m+1}_\Bob (\type^{\sun}_\Bob) & = %
1 \Big( 1 (\snature^{\sun}) , 1\big(\snature^{\sun} , 1(\snature^{\sun}) \big), %
\dots, %
\underbrace{1\big( 1 (\snature^{\sun}) , 1\big(\snature^{\sun} , 1(\snature^{\sun}) \big) \big), \dots}_{\text{$m+1$ times}}  \Big)  ,
\end{align*}
for every $m > 1$. Now, if we, as analysts, aim to capture the scenario truthfully, we may consider the state space $\States :=  \SNatures \times \Types_\Ann \times \Types_\Bob = \{ \snature^{\rain}, \snature^{\sun} \} \times \{ \type^{\rain}_\Ann \} \times \{\type^{\sun}_\Bob\}$, which is misaligned, as can be observed by focusing on the first two levels of the relevant CHBs. Namely,
\begin{align*}
\hierfun^{1}_\Ann (t^{\rain}_\Ann) & = 1 (\snature^{\rain}) , \\
\hierfun^{2}_\Ann (t^{\rain}_\Ann) & = 1 \big(\snature^{\rain} , 1(\snature^{\rain}) \big) ,
\end{align*}
for Ann and
\begin{align*}
\hierfun^{1}_\Bob (t^{\sun}_\Bob) & = 1 (\snature^{\sun}) , \\
\hierfun^{2}_\Bob (t^{\sun}_\Bob) & = 1 \big(\snature^{\sun} , 1(\snature^{\sun}) \big) .
\end{align*}
for Bob. In particular, misalignment as formalized in \Mref{def:misaligned_state_space} can be established by focusing, for example, on Ann, because
\begin{equation*}
(\projstfun^{2}_\Bob)^{-1} \big( \supp \marg_{\Hiers^{1}_{\Bob}} \hierfun^{2}_\Ann (\type^{\rain}_\Ann) \big) = %
(\projstfun^{2}_\Bob)^{-1} \big( 1(\snature^{\rain}) \big) \not\subseteq \Types_\Bob = \{\type^{\sun}_\Bob\}.
\end{equation*}
In fact, here we have the opposite (and strict) inclusion $\Types_\Bob \subsetneq (\projstfun^{2}_\Bob)^{-1} \big( 1(\snature^{\rain}) \big)  \subseteq \UTypes_\Bob$ as it is clear that there are multiple CHBs of Bob that assign probability $1$ to rain. We conclude by pointing out that we could have shown the misalignment of this state space also by focusing on Bob, again, with the problem arising at the $2^{\text{nd}}$-order belief of his CHB, specifically at $\hierfun^{2}_\Bob (\type^{\sun}_\Bob)$.
\end{example}

The example above demonstrates how the formalism we introduced can be applied to detect misalignment in a state space. However, it also highlights a significant limitation: checking for misalignment via CHBs is an extremely laborious process. Hence, there is a clear need for a more practical and systematic method to characterize---and operationalize---the potential misalignment of a state space, which is precisely what we aim to achieve next.

\begin{theorem}[State Space Characterization]
\label{th:state_space_characterization}
A state space $\States$ is misaligned if and only if it is non-belief-closed.
\end{theorem}

Thus, even if the original definition of misalignment is phrased in terms of hierarchies, with the additional merit that it makes the phenomenon elicitable, thanks to  \Mref{th:state_space_characterization} it is possible to take \emph{non-belief-closure} as a working `definition' of misalignment. In particular, we now exploit our characterization result by showing how \Mref{th:state_space_characterization} allows to handily capture misalignment by formalizing the narrative behind our recurring \Mref{ex:misalignment}.

\begin{example}[continues=ex:misalignment, name=Exploiting the State Space Characterization]
Recall that the actual state space of interest for us as analysts is 
\begin{equation*}
\States := %
\Sets { \state^1 , \state^2 } \equiv %
\SNatures \times \{\type^{\rain}_\Ann\} \times \{ \type^{\sun}_\Bob\} ,
\end{equation*}
keeping in mind that types \emph{only} stand for players' \emph{actual} CHBs. However, as previously noted, the state space $\States$ is not sufficiently \say{rich} to fully capture the complete hierarchy of beliefs (CHBs) of both agents. This limitation arises because $\States$ only accounts for their actual beliefs but does not allow to \emph{derive} these beliefs hierarchies based on reasoning within the state space. To address this issue, we consider a larger state space,  
\begin{equation*}
\widetilde{\States} := %
\Sets { \state^1 , \state^2 , \state^{\rain} , \state^{\sun} }
\end{equation*}
with
\begin{align*}
\state^1 & := (\snature^{\rain} , \type^{\rain}_\Ann , \type^{\sun}_\Bob ) , \\
\state^2 & := (\snature^{\sun} , \type^{\rain}_\Ann , \type^{\sun}_\Bob ) , \\
\state^{\rain} & := (\snature^{\rain} , \type^{\rain}_\Ann , \type^{\rain}_\Bob ) , \\
\state^{\sun} & := (\snature^{\sun} , \type^{\sun}_\Ann , \type^{\sun}_\Bob ),
\end{align*}
which incorporates additional states that allow this derivation. However, whereas $\widetilde{\States}$ provides a derivation of the belief hierarchies, we must be careful in how we interpret it. As analysts focusing on misalignment, our primary concern is not this enlarged state space \emph{per se}, but rather understanding which aspects of it reflect the actual beliefs of the players and which are artificial constructs introduced by our modeling choices. Thus, we should \emph{not} treat $\widetilde{\States}$ as our state space of interest. Indeed, the \emph{only} CHBs that capture what the players actually believe in the story described above, i.e., the \say{real} ones, are $\type^{\rain}_\Ann \in \UTypes_\Ann$ and $\type^{\sun}_\Bob \in \UTypes_\Bob$. It is evident, though, that the original state space $\States$ is \emph{not} belief-closed. To properly represent our original scenario, we \emph{only} need to have that
\begin{align*}
\belfun_{\Ann} (\type^{\rain}_{\Ann}) & = 1 (\snature^{\rain} , \type^{\rain}_{\Bob}) ,\\
\belfun_{\Bob} (\type^{\sun}_{\Bob}) & = 1 (\snature^{\sun} , \type^{\sun}_{\Ann}) .
\end{align*}
As a matter of fact, we are forced by the very definition of \say{standard} type structure to enrich $\States$ with $\state^{\rain}$ and $\state^{\sun}$ to derive the actual CHBs from the state space itself. These states are not part of the true underlying reality of the situation, but rather exist solely within the players’ reasoning. Specifically, $\state^{\rain}$ is an artifact of Ann’s perception of Bob’s reasoning, while $\state^{\sun}$ is an artifact of Bob’s perception of Ann’s reasoning and the corresponding types $\type^{\sun}_\Ann$ and $\type^{\rain}_\Bob$ should satisfy
\begin{align*}
\belfun_{\Ann} (\type^{\sun}_{\Ann}) & = 1 (\snature^{\sun} , \type^{\sun}_{\Bob})  ,\\
\belfun_{\Bob} (\type^{\rain}_{\Bob}) & = 1 (\snature^{\rain} , \type^{\rain}_{\Ann}) 
\end{align*}
to be consistent with our story. In particular, \Sref{fig:misalignment_example} provides a structured representation of the state spaces $\States$ and $\widetilde{\States}$, along with the players' beliefs (depicted via arrows), reinforcing the conceptual point we just made. As in previous illustrations, we highlight in red the players' CHBs that we---as outside omniscient observers---consider \say{real}, alongside the arrows representing the beliefs of the agents' types. However, in contrast to \Sref{fig:alignment_example}, we introduce an additional distinction: $\state^1$ and $\state^2$ are represented with filled dots to emphasize that these correspond to \say{real} states, whereas $\state^{\rain}$ and $\state^{\sun}$ are depicted with empty dots to highlight their \emph{imaginary} nature. These latter states do not \emph{actually} \say{exist} in an objective sense; rather, they exist only in the minds of the players. More specifically, the state $\state^{\rain}$ arises as a by-product of Ann's beliefs, while $\state^{\sun}$ results from Bob's. Consequently, the type $\type^{\rain}_\Bob$ does not \say{exist} in the sense that Bob does not actually hold the beliefs associated with this type---it is merely an artifact of Ann’s (incorrect) beliefs about Bob. The same reasoning applies to $\type^{\sun}_\Ann$, which emerges as a result of Bob’s beliefs about Ann. Ultimately, this misalignment highlights a fundamental issue: the players’ actual CHBs necessitate the introduction of additional states that, while required for a complete representation within \say{standard} type structures, do not correspond to any objectively real state that accurately captures the situation at hand.
\begin{figure}
\hspace{-2cm}
\begin{tikzpicture}
[real/.style={circle, draw, inner sep=1.5, fill=black},fake/.style={circle, draw, inner sep=1.5},scale=2] 
\node[real] (n_1) at (-0.5,0.5) {};
\node[real] (n_2) at (0.5,0.5) {};
\node[fake] (n_3) at (-0.5,-0.5) {};
\node[fake] (n_4) at (0.5,-0.5) {};
\draw (-0.7,0.7) node {$\state^1$}; 
\draw (0.7,0.7) node {$\state^2$};
\draw (-0.7,-0.7) node {$\state^{\rain}$};
\draw (0.7,-0.7) node {$\state^{\sun}$};
\draw[blue,->] (n_1) -- (n_3);
\draw[blue,->] (n_2) -- (n_3);
\path[blue,->] (n_3) edge  [loop left] node {} ();
\path[blue,dashed,->] (n_4) edge  [loop below] node {} ();
\draw[green,->] (n_1) -- (n_4);
\draw[green,->] (n_2) -- (n_4);
\path[green,->,dashed] (n_3) edge  [loop below] node {} ();
\path[green,->] (n_4) edge  [loop right] node {} ();
%
\matrix [below left] at (2.5,0.9) {
  \node [label=right:Ann] {}; \\
  \node [label=right:Bob] {}; \\
  \node [label=right:$\widetilde{\States}$] {}; \\
  \node [label=right:$\States$] {}; \\
};
\draw[blue] (1.7,0.7) -- (2,0.7);
\draw[green] (1.7,0.48) -- (2,0.48);
\draw[dashed] (1.7,0.2) -- (2,0.2);
\draw[] (1.7,-0.02) -- (2,-0.02);
\draw[rounded corners=7,dashed](-1,-1) rectangle (1,1);
\draw[rounded corners=7](-0.6,0.4) rectangle (0.6,0.6);
\draw (-1.2,0.85) node {\Large$\widetilde{\States}$};
\draw (0.75,0.35) node {\Large$\States$};
\draw (-3,0.7) node {$\state^1 := (\snature^\rain , \textcolor{red}{\type^{\rain}_\Ann} , \textcolor{red}{\type^{\sun}_\Bob} )$};
\draw (-3,0.4) node {$\state^2 := (\snature^\sun ,\textcolor{red}{\type^{\rain}_\Ann} , \textcolor{red}{\type^{\sun}_\Bob} )$};
\draw (-3,0.1) node {$\state^{\rain}  := (\snature^\rain , \textcolor{red}{\type^{\rain}_\Ann} , \type^{\rain}_\Bob )$};
\draw (-3,-0.2) node {$\state^{\sun} := (\snature^\sun , \type^{\sun}_\Ann , \textcolor{red}{\type^{\sun}_\Bob})$};
\end{tikzpicture}
\caption{A representation of the agents' beliefs in $\widetilde{\States}$.}
\label{fig:misalignment_example}
\end{figure}
\end{example}

\section{A Framework for Misalignment}
\label{sec:a_framework_for_misalignment}

Having formally defined and characterized misalignment in \Mref{th:state_space_characterization}, we now aim to conduct interactive epistemological analyses to derive behavioral predictions, even in the presence of misalignment. However, this goal presents an immediate challenge because---according to \Mref{th:state_space_characterization}---addressing misalignment requires working with non-belief-closed state spaces. The inherent nature of such spaces makes interactive epistemological reasoning not just difficult but, in many cases, highly impractical. This naturally leads to the following question: can we transform a misaligned state space into a belief-closed one? Achieving this would allow us to retain the analytical advantages of belief closure while still accounting for misalignment.

However, this transformation cannot be carried out in a na\"{i}ve way by simply adding the missing types to the misaligned state space of interest, as doing so would effectively recreate the issues depicted in our running example in \Sref{sec:background_knowledge}. In other words, extra care must be taken when introducing the additional types needed to ensure that the CHBs of interest are derivable within a belief-closed state space. To address this, we introduce in \Sref{subsec:agent-dependent_type_structures} the notion of an \emph{agent-dependent type structure}, which demonstrates that accounting for misalignment reveals hidden assumptions embedded in the \say{standard} type structures and state spaces we typically work with.

In \Sref{subsec:agent-dependent_type_structures}, we treat agent-dependent type structures as given and explore their implications, focusing on what they allow us to achieve rather than on how they are constructed. The construction itself is provided in \Sref{subsec:the_agent_closure}, where we develop a method for obtaining a special kind of agent-dependent type structure, namely, the \emph{minimal} agent-dependent type structure with respect to our CHBs of interest. This distinction allows us to separate the conceptual use of agent-dependent type structures from their formal foundation. Due to the modular structure of these two sections, a reader interested in applying agent-dependent type structures without concern for their underlying construction can focus on \Sref{subsec:agent-dependent_type_structures} and skip \Sref{subsec:the_agent_closure}.

\subsection{Agent-Dependent Type Structures}
\label{subsec:agent-dependent_type_structures}

To account for the possibility that agents may reason differently from us as analysts---given what we consider possible---we introduce a definition of a \say{standard} type structure that can vary across individual agents. This allows us to formalize different reasoning, but employing the usual machinery.

\begin{definition}[Agent-Dependent Type Structure]
\label{def:agent-dependent_type_structure}
Given a state space $\States :=  \SNatures \times \prod_{j \in \Agents} \Types_j$ and an agent $i \in \Agents$, an \emph{agent-$i$-dependent type structure induced by $\Types_i$} is a \say{standard} type structure
\begin{equation*}
\ADTypeSt^i (\Types_i) := 
\Angles { \Agents, \SNatures , \Rounds {\ADTypes^{i}_j , \ADbelfun^{i}_j }_{j \in \Agents} } ,
\end{equation*}
such that, for every $j \in \Agents$,
\begin{enumerate}[leftmargin=*,label=\arabic*)]
\item the set $\ADTypes^{i}_j \subseteq  \UTypes_j$ is agent $j$'s compact \emph{agent-$i$-dependent type space}.

\item the function $\ADbelfun^i_j : \ADTypes^{i}_j  \to \Delta (\SNatures \times \ADTypes^{i}_{-j})$ is agent $j$'s \emph{agent-$i$-dependent (continuous) belief function} given by $\ADbelfun^i_j:=\belhom_{j}\vert_{\ADTypes^{i}_j} $.
\end{enumerate}
with
\begin{enumerate}[leftmargin=*,label=\arabic*)]
\setcounter{enumi}{2}
\item $\Types_i \subseteq \ADTypes^{i}_i$.
\end{enumerate}
The set $\Types_i$ is the set of \emph{real types} of agent $i$, whereas $\ADTypes^{i}_i \setminus \Types_i$ is the set of \emph{imaginary types}.
\end{definition}

In what follows, we let $\FamADTypeSt^i (\Types_i)$ denote the collection of agent-$i$-dependent type structures. Now, it has to be observed that the universal type structure is always an agent-$i$-dependent type structure for any player $i$ and actually also for any given state space. Thus, $\FamADTypeSt^i (\Types_i)$ is nonempty. On the contrary, it is often the case that there are multiple elements in $\FamADTypeSt^i (\Types_i)$ as illustrated with our running example below.

Given a state space $\States :=  \SNatures \times \prod_{j \in \Agents} \Types_j$, an agent $i \in \Agents$, and an agent-$i$-dependent type structure $\ADTypeSt^i (\Types_i)$, we define the corresponding \emph{agent-$i$-dependent state space}, denoted by $\ADStates^i$, as the belief-closed state space induced by $\ADTypeSt^i (\Types_i)$ in light of \Mref{rem:belief-closed_type_structure}. 

The distinction between real and imaginary types leads to an important structural property of agent-dependent type structures. We say that $\ADTypeSt^i (\Types_i)$ is \emph{degenerate} if $\ADStates^i \setminus \States = \emptyset$, i.e., the belief-closed state space does not \say{introduce} any new states beyond those in $\States$. Conversely, if $\ADStates^i$ contains additional states beyond $\States$, the type structure is \emph{non-degenerate}. This classification naturally extends to profiles of agent-dependent type structures.

\begin{definition}[Degenerate Profile of Agent-Dependent Type Structures]
\label{def:degenerate_agent-dependent_type_structure}
Given a state space $\States := \SNatures \times \prod_{j \in \Agents}  \Types_j$, a profile of agent-dependent type structures $\big( \ADTypeSt^j (\Types_j) \big)_{j \in \Agents}$ is \emph{degenerate}
if every agent-$i$-dependent type structure $\ADTypeSt^i (\Types_i)$ is degenerate. Otherwise, the profile is \emph{non-degenerate}.\footnote{That is, a profile is non-degenerate if there exists at least one agent $i \in \Agents$ such that $\ADTypeSt^i (\Types_i)$ is non-degenerate.}
\end{definition}

\begin{definition}[Common Profile of Agent-Dependent Type Structures]
\label{def:common_agent-dependent_type_structure}
Given a state space $\States :=  \SNatures \times \prod_{j \in \Agents} \Types_j$, a profile of agent-dependent type structures $\big( \ADTypeSt^j (\Types_j) \big)_{j \in \Agents}$ is \emph{common} if
$
\ADStates^i = \ADStates^j
$
for every $i, j \in \Agents$. Otherwise, the profile is \emph{non-common}.
\end{definition}

By combining these two classifications, we obtain a taxonomy of profiles of agent-dependent type structures, summarized in \Sref{tab:misalignment_taxonomy}. A key observation is that a degenerate profile must necessarily be common. That is, if no new states are introduced for any agent, then all agents must share the same belief-closed state space. This fact, explicitly shown in \Sref{tab:misalignment_taxonomy} via an $\times$, is formalized in the following remark.

\begin{table}
\begin{center}
\renewcommand{\arraystretch}{1.2}
\centering
\begin{tabular}{|c|c|c|} 
\hline
					& Common 	& Non-Common \\ 
\hline
\rule{0pt}{3ex}
Degenerate     		&   	\say{Standard} State Spaces			&  $\times$   \\
Non-Degenerate 	&  $\circ$				&    $\circ$   \\ 
\hline
\end{tabular}
\end{center}
\caption{A taxonomy of state spaces given a profile of agent-dependent type structures.}
\label{tab:misalignment_taxonomy}
\end{table}

\begin{remark}
\label{rem:degenerate_common}
Given a state space $\States :=  \SNatures \times \prod_{j \in \Agents}\Types_j$, if a profile of agent-dependent type structures $\big( \ADTypeSt^j (\Types_j) \big)_{j \in \Agents}$ is degenerate, then it must be common.
\end{remark}

The taxonomy in \Sref{tab:misalignment_taxonomy}, along with \Mref{rem:degenerate_common}, highlights a fundamental feature of agent-dependent type structures: the conventional approach of assuming a single---common---type structure inherently presupposes a degenerate profile. Recognizing this assumption is not just a conceptual refinement, but it has significant implications. As we show in \Sref{sec:misalignment_matters}, allowing for non-degenerate and non-common profiles leads to \emph{novel} behavioral predictions that do not arise under traditional epistemic frameworks. To make these ideas concrete, we now illustrate how agent-dependent type structures operate in our leading example.

\begin{example}[continues=ex:misalignment, name=Taking Care of Misalignment with Agent-Dependent Type Structures]
As before, we consider the state space  
\begin{equation*}
\States := %
\Sets { \state^1 , \state^2 } \equiv %
\SNatures \times \{\type^{\rain}_\Ann\} \times \{ \type^{\sun}_\Bob\} ,
\end{equation*}
which represents the actual state space of interest from the perspective of an analyst. For Ann, we define the agent-dependent type structure as  
\begin{equation*}
\ADTypeSt^\Ann (\Types_\Ann) := 
\Angles { \Agents, \SNatures , \Rounds {\ADTypes^{\Ann}_j , \ADbelfun^{\Ann}_j }_{j \in \Agents} } ,
\end{equation*}
where the type spaces are given by  
\begin{align*}
\ADTypes^{\Ann}_{\Ann}  := \Sets { \type^{\rain}_{\Ann} , \type^{\frac{1}{2}}_\Ann } , &&
&\ADTypes^{\Ann}_{\Bob}  := \Sets { \type^{\rain}_{\Bob} , \type^{\frac{1}{2}}_\Bob }. 
\end{align*}
The corresponding belief functions are  
\begin{align*}
\ADbelfun^{\Ann}_{\Ann} (\type^{\rain}_{\Ann}) & := 1 (\snature^\rain, \type^{\rain}_{\Bob}) , &&
&\ADbelfun^{\Ann}_{\Ann} \Rounds{\type^{\frac{1}{2}}_{\Ann}} & := \Rounds{ \frac{1}{2} (\snature^\rain, \type^{\rain}_{\Bob}) , \frac{1}{2} \Rounds {\snature^\rain, \type^{\frac{1}{2}}_{\Bob}} }, \\
\ADbelfun^{\Ann}_{\Bob} (\type^{\rain}_{\Bob}) & := 1 (\snature^\rain, \type^{\rain}_{\Ann}) , &\text{and}&
&\ADbelfun^{\Ann}_{\Bob} \Rounds{\type^{\frac{1}{2}}_{\Bob}} & :=  \Rounds{ \frac{1}{2} (\snature^\rain, \type^{\rain}_{\Ann}) , \frac{1}{2} \Rounds{\snature^\rain, \type^{\frac{1}{2}}_{\Ann}} }. 
\end{align*}
Similarly, for Bob, we define  
\begin{equation*}
\ADTypeSt^\Bob (\Types_\Bob) := 
\Angles { \Agents, \SNatures , \Rounds {\ADTypes^{\Bob}_j , \ADbelfun^{\Bob}_j }_{j \in \Agents} } ,
\end{equation*}
with type spaces  
\begin{align*}
\ADTypes^{\Bob}_{\Bob} & := \Sets { \type^{\sun}_{\Bob} , \type^{\frac{1}{3}}_{\Bob} } , &&
&\ADTypes^{\Bob}_{\Ann} & := \Sets { \type^{\sun}_{\Ann} , \type^{\frac{2}{3}}_\Ann }. 
\end{align*}
The belief functions in Bob’s type structure are  
\begin{align*}
\ADbelfun^{\Bob}_{\Bob} (\type^{\sun}_{\Bob}) & := 1 (\snature^\sun, \type^{\sun}_{\Ann}) , &&
& \ADbelfun^{\Bob}_{\Bob} \Rounds{\type^{\frac{1}{3}}_{\Bob}} & :=  \Rounds{ \frac{1}{3} (\snature^\sun, \type^{\sun}_{\Ann}) , \frac{2}{3} \Rounds{\snature^\sun, \type^{\frac{2}{3}}_{\Ann}} }, \\
\ADbelfun^{\Bob}_{\Ann} (\type^{\sun}_{\Ann}) & := 1 (\snature^\sun, \type^{\sun}_{\Bob}) , &\text{and}&
&\ADbelfun^{\Bob}_{\Ann} \Rounds{\type^{\frac{2}{3}}_{\Ann}} & :=  \Rounds{ \frac{2}{3} (\snature^\sun,  \type^{\sun}_{\Bob}) , \frac{1}{3} \Rounds{\snature^\sun, \type^{\frac{1}{3}}_{\Bob}} }. 
\end{align*}
By construction, this profile of agent-dependent type structures is both non-common and non-degenerate.
\end{example}

One important point must be emphasized regarding the example just discussed. The agent-dependent type structures introduced there are not \say{minimal} with respect to the original CHBs of interest from the analyst’s perspective. To see this, consider Ann’s agent-dependent type structure. The type $\ADtype^{\frac{1}{2}}_\Ann$ is not, strictly speaking, necessary to construct an agent-$\Ann$-dependent type structure that includes $\ADtype^{\rain}_\Ann$. In fact, the only essential element for this purpose is $\ADtype^{\rain}_\Bob$. This does not imply a conceptual flaw in the construction, but instead it highlights a potential concern regarding \emph{robustness} in the analyst’s modeling choices. However, although incorporating additional types may sometimes be justifiable, it is also important to identify agent-dependent type structures that satisfy a well-defined minimality condition aligned with the desiderata outlined above, which is what we address in the next subsection by formally defining and constructing such minimal structures.

\subsection{The Agent Closure}
\label{subsec:the_agent_closure}

In this section, we focus on minimal agent-dependent type structures. We begin by defining them to then establish their existence through a constructive argument. In doing so, we provide a formal foundation for the broader concept of agent-dependent type structures.

To this end, we introduce the definition of a minimal agent-dependent type structure, which can be understood as the minimal (in a precise sense to be introduced) \say{standard} type structure that qualifies as an agent-dependent type structure while containing only the following:  
\begin{enumerate}[leftmargin=*,label=$\bullet$]
\item the types required to construct the CHBs of interest, as determined by the original state space under consideration, and  

\item the types necessary to ensure that the structure remains a well-defined \say{standard} type structure from a topological standpoint.  
\end{enumerate}
The second condition follows from the requirement that type spaces in a \say{standard} type structure must be compact and nonempty, as stipulated in \Sref{sec:background_knowledge}, and therefore also shows up in \Mref{def:agent-dependent_type_structure}.

\begin{definition}[Minimal Agent-Dependent Type Structure]
\label{def:minimal_agent-dependent_type_structure}
Given a state space $\States := \prod_{j \in \Agents} \SNatures \times \Types_j$ and an agent $i \in \Agents$, an agent-$i$-dependent type structure 
\begin{equation*}
\ADTypeSt^i (\Types_i) := 
\Angles { \Agents, \SNatures , \Rounds {\ADTypes^{i}_j , \ADbelfun^{i}_j }_{j \in \Agents} } \in \FamADTypeSt^i (\Types_i),
\end{equation*}
is \emph{minimal} if it is minimal with respect to set inclusion within $\FamADTypeSt^i (\Types_i)$, that is, if $\ADTypeSt^i (\Types_i) \subseteq \ADTypeSt^{\prime,i} (\Types_i)$ for every $\ADTypeSt^{\prime,i} (\Types_i) \in \FamADTypeSt^i (\Types_i)$.
\end{definition}

Anticipating the main result of this section, we let 
\begin{equation*}
\minADTypeSt^{i} (\Types_i) := 
\Angles { \Agents, \SNatures , \Rounds {\minADTypes^{i}_j , \minADbelfun^{i}_j }_{j \in \Agents} } ,
\end{equation*}
denote the minimal agent-$i$-dependent type structure, with $\minADStates^{i}$ denoting the corresponding minimal agent-$i$-dependent state space. Furthermore, a profile of agent-dependent type structures $\big( \ADTypeSt^j (\Types_j) \big)_{j \in \Agents}$ is \emph{minimal} if $\ADTypeSt^i (\Types_i)$ is minimal for every $i \in \Agents$, i.e., if $\ADTypeSt^i (\Types_i) = \minADTypeSt^{i} (\Types_i)$ for every $i \in \Agents$.

\begin{example}[continues=ex:misalignment, name=Taking Care of Misalignment with a Profile of Minimal Agent-Dependent Type Structures]
As before, we have 
\begin{equation*}
\States := %
\Sets { \state^1 , \state^2 } \equiv %
\SNatures \times \{\type^{\rain}_\Ann\} \times \{ \type^{\sun}_\Bob\} ,
\end{equation*}
as the actual state space of interest for us as analysts. Now, it is immediate to observe that regarding Ann we have 
\begin{equation*}
\minADTypeSt^{\Ann} (\Types_\Ann) := 
\Angles { \Agents, \SNatures , \Rounds {\minADTypes^{\Ann}_j , \minADbelfun^{\Ann}_j }_{j \in \Agents} } ,
\end{equation*}
with  
\begin{align*}
\minADTypes^{\Ann}_{\Ann} & := \{ \type^{\rain}_{\Ann} \} ,\\
\minADTypes^{\Ann}_{\Bob} & := \{ \type^{\rain}_{\Bob} \} ,
\end{align*}
and
\begin{align*}
\ADbelfun^{\Ann}_{\Ann} (\type^{\rain}_{\Ann}) & := 1 (\snature^\rain, \type^{\rain}_{\Bob}) ,\\
\ADbelfun^{\Ann}_{\Bob} (\type^{\rain}_{\Bob}) & := 1 (\snature^\rain, \type^{\rain}_{\Ann}) ,
\end{align*}
such that $\minADStates^\Ann := \{\state^{\rain}\}$. Also, regarding Bob we have
\begin{equation*}
\minADTypeSt^{\Bob} (\Types_\Bob) := 
\Angles { \Agents, \SNatures , \Rounds {\minADTypes^{\Bob}_j , \minADbelfun^{\Bob}_j }_{j \in \Agents} } ,
\end{equation*}
with  
\begin{align*}
\minADTypes^{\Bob}_{\Bob} & := \{ \type^{\sun}_{\Bob} \} ,\\
\minADTypes^{\Bob}_{\Ann} & := \{ \type^{\sun}_{\Ann} \} ,
\end{align*}
and
\begin{align*}
\ADbelfun^{\Bob}_{\Bob} (\type^{\sun}_{\Bob}) & := 1 (\snature^\sun, \type^{\sun}_{\Ann}) ,\\
\ADbelfun^{\Bob}_{\Ann} (\type^{\sun}_{\Ann}) & := 1 (\snature^\sun, \type^{\sun}_{\Bob}) ,
\end{align*}
such that $\minADStates^\Bob := \{\state^{\sun}\}$. At least intuitively, it should be clear that the profile of agent-dependent type structures is indeed minimal, \emph{and} non-common and non-degenerate.
\end{example}

Having introduced the notion of minimal agent-dependent type structure, a rather natural question to ask is if this object always exists. Anf if so, if there is a more constructive description of this object. Thus, to answer these questions, given a state space $\States := \SNatures \times T$ and an arbitrary agent $i \in \Agents$, we let
\begin{equation}
\label{eq:closure_operator}
\Closure_i (\States) := \SNatures \times \Types_i \times \overline{\Rounds{\bigcup_{\type_i \in \Types_i} \supp \typehom_i (\type_i) }} 
\end{equation}
denote the \emph{closure operator of agent} $i$ that we use to define the following notion of closure of a state space from the perspective of an agent $i$.\footnote{A similar idea is hinted in \citet[Chapter III.1.b(5), p.137]{Mertens_et_al_2015}. See also \citet{Gossner_Veiel_2024}.}

\begin{definition}[Agent Closure with Respect to a State Space]
\label{def:agent_closure}
Given a state space $\States := \prod_{j \in \Agents} \SNatures \times \Types_j$ and an agent $i \in \Agents$, the \emph{agent closure with respect to $\States$ of $i$} is the state space 
\begin{equation}
\label{eq:agent_closure}
\ac_i (\States) := %
\overline{\Rounds{ \bigcup_{\ell \in \Naturals} \widehat{\States}^\ell }} ,\footnote{Here, by employing standard notation, given a topological space $\ArbSet$, we let $\overline{\ArbSet}$ denote its (topological) closure.}
\end{equation}
inductively defined on $\Naturals$ as:
\begin{itemize}[leftmargin=*]
\item {$[n =1]$} $\widehat{\States}^1  := \States \cup \Closure_i (\States)$;

\item {$[n \geq 1]$} for every $n \in \Naturals$, assuming that $\widehat{\States}^{n}$ has been defined, we let
\begin{equation}
\label{eq:agent_closure_inductive}
\widehat{\States}^{n+1} := %
\widehat{\States}^{n} \cup \bigcup_{j \in \Agents} \Closure_j (\widehat{\States}^n) .
\end{equation}
\end{itemize}
\end{definition}

Letting $\ac_i (T_j)$ denote the $j$-component of $\ac_i (\States)$ for every $j \in \Agents$, we have 
\begin{equation*}
\ac_i (\States) = \SNatures \times \prod_{j \in \Agents} \ac_i (T_j) .
\end{equation*}
With an abuse of the terminology (although innocent by the result that follows), the set $\Types_i$ is the set of \emph{real types} of agent $i$, while the set of types $\ac_i (\Types_i) \setminus \Types_i$ is the set of \emph{imaginary types} of agent $i$. Moreover, an agent closure $\ac_i (\States)$ is \emph{degenerate} if $\ac_i (\States) \setminus \States = \emptyset$ and is \emph{non-degenerate} otherwise. Finally, from \Mref{rem:belief-closed_type_structure}, we let $\ADTypeSt^{i} (\ac_i (\States))$ denote agent-$i$-dependent type structure induced by $\ac_i (\States)$.

Now, we are ready to state the main result of this section, namely, that minimal agent-dependent type structures and agent closures are essentially the same, they exist, are unique, and are belief-closed. The result essentially follows from a sort of usual dual argument in order theory, which we present in detail in \Sref{subapp:proof_a_framework_for_misalignment}.

\begin{theorem}
\label{th:characterization_minimality}
Fix a state space $\States := \prod_{j \in \Agents} \SNatures \times \Types_j$. For every $i \in \Agents$,
\begin{enumerate}[leftmargin=*,label=\arabic*)]
    \item $\ac_i (\States)$ is compact (hence, measurable) and belief-closed,
    \item $\ac_i (\States)$  and $\minADTypeSt^{i} (\Types_i)$ are nonempty,
    \item $\ac_i (\States)$  and $\minADTypeSt^{i} (\Types_i)$ are unique,
\end{enumerate}
and
\begin{align*}
\minADTypeSt^{i} (\Types_i) & = \ADTypeSt^{i} (\ac_i (\States)) ,\\
\minADStates^i & = \ac_i (\States) .
\end{align*}
\end{theorem}

Having provided a foundation for the notion of agent-dependent type structure via the notion of agent closure, we now see how the agent closure works in our example.

\begin{example}[continues=ex:misalignment, name=Taking Care of Misalignment with the Agent Closure]
As before, we have 
\begin{equation*}
\States := %
\Sets { \state^1 , \state^2 } \equiv %
\SNatures \times \{\type^{\rain}_\Ann\} \times \{ \type^{\sun}_\Bob\} ,
\end{equation*}
as the actual state space of interest for us as analysts. Now, it is immediate to observe that we have 
\begin{align*}
\ac_\Ann (\States) & := \{\snature^{\rain}\} \times \{t^{\rain}_\Ann\} \times \{t^{\rain}_\Bob\} \hspace{0.9mm} \equiv \{ \state^{\rain} \} \\
\ac_\Bob (\States) & := \{\snature^{\sun}\} \times \{\type^{\sun}_\Ann\} \times \{\type^{\sun}_\Bob\} \equiv \{ \state^{\sun} \} ,
\end{align*}
from which we define for every $i \in \Agents$ the corresponding agent-dependent type structure $\ADTypeSt^i (\ac_i (\States))$. In particular, regarding Ann we have
\begin{align*}
\ac_\Ann (\Types_\Ann) & := \{ \type^{\rain}_{\Ann} \} ,\\
\ac_\Ann (\Types_\Bob) & := \{ \type^{\rain}_{\Bob} \} ,
\end{align*}
with 
\begin{align*}
\acADbelfun^{\Ann}_{\Ann} (\type^{\rain}_{\Ann}) & := 1 (\snature^\rain, \type^{\rain}_{\Bob}) ,\\
\acADbelfun^{\Ann}_{\Bob} (\type^{\rain}_{\Bob}) & := 1 (\snature^\rain, \type^{\rain}_{\Ann}) ,
\end{align*}
while regarding Bob we have 
\begin{align*}
\ac_\Bob (\Types_\Bob) & := \{ \type^{\sun}_{\Bob} \} ,\\
\ac_\Bob (\Types_\Ann) & := \{ \type^{\sun}_{\Ann} \} .
\end{align*}
with
\begin{align*}
\acADbelfun^{\Bob}_{\Bob} (\type^{\sun}_{\Bob}) & := 1 (\snature^\sun, \type^{\sun}_{\Ann}) ,\\
\acADbelfun^{\Bob}_{\Ann} (\type^{\sun}_{\Ann}) & := 1 (\snature^\sun, \type^{\sun}_{\Bob}) .
\end{align*}
Once more, it is immediate to observe that the profile of agent-dependent type structures induced by the profile $\big( \ac_j (\States) \big)_{j \in \Agents}$ is non-common and non-degenerate.
\end{example}

\section{Interactive Epistemology under Misalignment}
\label{sec:interactive_epistemology_under_misalignment}

Having introduced in the previous section an opportune framework to take care of the potential presence of misalignment, the latter being a distinct feature of interactive environments, we now show how interactive epistemology can be performed in presence of misalignment as captured via agent-dependent type structures. 

For this purpose, first of all, we recall some basic facts about the \say{standard} framework. In particular, given a \say{standard} type structure
\begin{equation*}
\TypeSt := %
\Angles { \Agents, \SNatures , \Rounds { \belfun_j , \Types_j}_{j \in \Agents} }
\end{equation*}
and an arbitrary agent $i \in \Agents$, an \emph{agent's $i$ event} is a closed product subset $\Event_{-i} \subseteq \SNatures \times \prod_{j \neq i}\Types_{j}$. Thus, the \emph{belief operator} $\Bel_{i}$ of agent $i$ is defined as 
\begin{equation*}
\Bel_{i} (\Event_{-i}) := \Set { (\snature, \type_i) \in \SNatures \times \Types_i |  \belfun_{i} (\type_i) (\Event_{-i}) = 1  },
\end{equation*}
for every agent $i$'s event $\Event_{-i}$. It is well known that the belief operator satisfies the following properties. 

\begin{remark}
\label{eq:belief_operator_properties}
Given a state space $\States := \SNatures \times \prod_{j \in \Agents} \Types_{j}$ and an agent $i \in \Agents$:
\begin{enumerate}[leftmargin=*,label=\arabic*)]
\item (Necessitation) $\Bel_i (\SNatures \times \Types_{-i}) = \Types_{i}$;

\item (Monotonicity) for every agent's $i$ events $\Event_{-i}, \Event'_{-i}$, if $\Event_{-i} \subseteq \Event'_{-i}$, then $\Bel_i (\Event_{-i}) \subseteq \Bel_i (\Event'_{-i})$;

\item (Conjunction) for every  agent's $i$ events $\Event_{-i}, \Event'_{-i}$, $\Bel_i (\Event_{-i}) \cap \Bel_i (\Event'_{-i}) = \Bel_i (\Event_{-i} \cap \Event'_{-i})$;

\item (Positive Introspection) for every  agent's $i$ event $\Event_{-i}$, $\Bel_i (\Bel_i (\Event_{-i})) = \Bel_i (\Event_{-i})$;

\item (Negative Introspection) for every  agent's $i$ event $\Event_{-i}$, $\Bel_i \big( [\Bel_i (\Event_{-i} )]^\complement \big) = [\Bel_i (\Event_{-i})]^\complement$.\footnote{As it is customary, given an arbitrary set $\ArbSet$, we let $\ArbSet^\complement$ denote its complement.}
\end{enumerate}
\end{remark}

Letting an \emph{event} $\Event$ be a closed product subset of $\States := \SNatures \times \prod_{j \in \Agents} \Types_{j}$, i.e., $\Event \subseteq \SNatures \times \prod_{j \in \Agents} \Types_{j}$, from the belief operator, we define the \emph{mutual belief operator} as $\Bel (\Event) := \prod_{j \in \Agents} \Bel_j (\Event_{-j})$, for every event $\Event$. 

Finally, we define the \emph{common correct belief} operator $\CB (\Event) := E \cap \Bel (\Event)$ and, building on it for the purpose of capturing interactive reasoning, we define inductively the event that captures \emph{event $\Event$ and $m$-correct belief in event $\Event$} (henceforth, EmBE) by setting $\CB^0 (\Event) = \Event$ and 
\begin{equation*}
\CB^{m} (\Event)  :=  \Event \cap \bigcap^{m - 1}_{k = 0} \Bel (\CB^{k} (\Event)) ,
\end{equation*}
for every $m \in \Naturals$, with
\begin{equation*}
\label{eq:ECBE}
\CB^{\infty} (\Event)  :=  \bigcap_{\ell \geq 0} \CB^\ell (\Event) 
\end{equation*}
denoting the event capturing \emph{event $\Event$ and common correct belief in event $\Event$} (henceforth, ECBE), where all the previous events are closed and---as a consequence---measurable in our framework by standard arguments. 

Now, given a state space $\States := \prod_{j \in \Agents} \SNatures \times \Types_j$, an agent $i \in \Agents$, and an agent-dependent type structure $\ADTypeSt^i (\Types_i)$  with corresponding agent-dependent state space $\ADStates^{i}$, we introduce the operators defined above opportunely modified to address the presence of misalignment. In particular, for every agent $i$'s agent-dependent type structure $\ADTypeSt^i (\Types_i)$ and for every agent $i$'s event $\ADEvent^{i}_{-i} \subseteq \SNatures \times \prod_{j \neq i}\ADTypes^{i}_{j}$, we let 
\begin{equation*}
\ADBel_{i} (\ADEvent^{i}_{-i}) := \Bel_i (\ADEvent^{i}_{-i}) \cap \Types_i,
\end{equation*}
with $\Bel_i$ employed on $\ADStates^i$, whereas for every event $\ADEvent^{i} \subseteq \SNatures \times \prod_{i \in \Agents}\ADTypes^i$, we let
\begin{equation}
\label{eq:agent-dependent_CB}
\ADCB[m]_i (\ADEvent^i) := %
\Rounds { \proj_{\ADTypes^{i}_i} \CB^{m} (\ADEvent^i) } \cap \Types_i
\end{equation}
denote the event that captures \emph{agent $i$'s real EmBE}, for every $m \in \Naturals$. Thus, crucially, first the analysis is performed via the \say{standard} operators $\Bel_i$ or $\CB^m$ on the state space $\ADStates^i$ \emph{and then} the resulting $\Bel_{i} (\ADEvent^i_{-i})$ or $\CB^{m} (\ADEvent^i)$ are intersected with the real types in $\Types_i$, where the intersection is possibly empty.

Building on the last paragraph, we have 
\begin{equation}
\label{eq:player_real_ECBE}
\ADCB[\infty]_i (\ADEvent^i) := \bigcap_{\ell \geq 0} \ADCB[\ell]_i (\ADEvent^i)
\end{equation}
denoting the event capturing \emph{agent $i$'s real ECBE}, where it is important to  observe that, even if all these definitions crucially depend on the agent-dependent type structure $\ADTypeSt^i (\Types_i)$ of agent $i$, we keep this dependence notationally implicit.

Finally, given a state space $\States$, a profile of agent-dependent type structures $\big( \ADTypeSt^j (\Types_j ) \big)_{j \in \Agents}$, and an event $\ADEvent^\bullet := (\ADEvent^{j})_{j \in \Agents} \subseteq \prod_{j \in \Agents} (\SNatures \times \prod_{k \in \Agents}\ADTypes_k^j)$, we let $\ADCB[m] (\ADEvent^\bullet) :=  \prod_{j \in \Agents} \ADCB[m]_j (\ADEvent^j)$ denote the event that captures \emph{real EmBE}, for every $m \in \Naturals$, with
\begin{equation}
\label{eq:real_CB}
\ADCB[\infty] (\ADEvent^\bullet) :=  \prod_{j \in \Agents} \ADCB[\infty]_j (\ADEvent^j)
\end{equation}
denoting the event that captures \emph{real ECBE}.

\begin{example}[continues=ex:misalignment, name=Interactive Epistemology with Agent-Dependent Type Structures]
Focusing on the profile of non-common and non-degenerate minimal agent-dependent type structures $(\minADTypeSt^i (\Types_j)\big)_{j \in \Agents}$ with corresponding profile of agent-dependent state spaces $\big( \minADStates^j \big)_{j \in \Agents}$, for Ann we have
\begin{align*}
\minADTypes^{\Ann}_{\Ann} & := \{ \type^{\rain}_{\Ann} \} ,\\
\minADTypes^{\Ann}_{\Bob} & := \{ \type^{\rain}_{\Bob} \} ,
\end{align*}
with 
\begin{align*}
\ADbelfun^{\Ann}_{\Ann} (\type^{\rain}_{\Ann}) & := 1 (\snature^\rain, \type^{\rain}_{\Bob}) ,\\
\ADbelfun^{\Ann}_{\Bob} (\type^{\rain}_{\Bob}) & := 1 (\snature^\rain, \type^{\rain}_{\Ann}) ,
\end{align*}
such that $\minADStates^\Ann := \{ \state^{\rain} \}$, while regarding Bob we have 
\begin{align*}
\minADTypes^{\Bob}_{\Ann} & := \{ \type^{\sun}_{\Ann} \} ,\\
\minADTypes^{\Bob}_{\Bob} & := \{ \type^{\sun}_{\Bob} \} .
\end{align*}
with
\begin{align*}
\ADbelfun^{\Bob}_{\Ann} (\type^{\sun}_{\Ann}) & := 1 (\snature^\sun, \type^{\sun}_{\Bob}) ,\\
\ADbelfun^{\Bob}_{\Bob} (\type^{\sun}_{\Bob}) & := 1 (\snature^\sun, \type^{\sun}_{\Ann}) . 
\end{align*}
such that $\minADStates^\Bob := \{ \state^{\sun} \}$. Now, assuming that the event of interest is $\{\state^r, \state^n\}$ in $\widetilde{\States}$, we have 
\begin{equation*}
\CB^\infty (\{\state^{\rain}, \state^{\sun}\}) = %
\{%
(\snature^\rain,\type^{\rain}_{\Ann},\type^{\rain}_\Bob ), %
(\snature^\sun, \type^{\sun}_{\Ann}, \type^{\sun}_\Bob )
\}
\end{equation*}
However, being attentive to misalignment, we focus on agent-dependent type structures and on the corresponding event $\{\state^{\rain}, \state^{\sun}\}^{\bullet} = \big(\{\state^{\rain}, \state^{\sun}\}^\Ann , \{\state^{\rain}, \state^{\sun}\}^\Bob\big)$. Thus, we obtain
\begin{equation*}
\CB^\infty (\{\state^{\rain}, \state^{\sun}\}^\Ann) = \{(\snature^\rain, \type^{\rain}_{\Ann}, \type^{\rain}_\Bob )\}
\end{equation*}
in $\minADStates^\Ann$ and
\begin{equation*}
\CB^\infty (\{\state^{\rain}, \state^{\sun}\}^{\Bob}) = \{(\snature^\sun, \type^{\sun}_\Ann, \type^{\sun}_\Bob )\} 
\end{equation*}
in  $\minADStates^\Bob$ with
\begin{align*}
\ADCB[\infty]_\Ann (\{\state^{\rain}, \state^{\sun}\}^\Ann) & = \{ \type^{\rain}_\Ann \} ,\\
\ADCB[\infty]_\Bob (\{\state^{\rain}, \state^{\sun}\}^\Bob) & = \{ \type^{\sun}_\Bob \}. \qedhere
\end{align*}
\end{example}

The following---crucial---result regarding the measurability of all the objects defined in this section is an immediate consequence of the closedness of the belief and the common correct belief (with its iterated application) operators starting from closed events and the closedness of countable intersections.

\begin{proposition}[Measurability]
\label{prop:real_CB_measurability}
Given a state space $\States$, an agent $i \in \Agents$, and an agent-dependent type structure $\ADTypeSt^i (\Types_i)$, for every event $\ADEvent^i$ and $m \in \Naturals$, the events $\ADCB[m]_i (\ADEvent^i)$, $\ADCB[m] (\ADEvent^i)$, $\ADCB[\infty]_i (\ADEvent^i)$, and $\ADCB[\infty]_i (\ADEvent^i)$ are closed, hence, measurable.
\end{proposition}

We can now enunciate the following collection of results, which are remarks stated as propositions in light of the fact that, even if mathematically trivial, they happen to be extremely important from a conceptual standpoint.

\begin{proposition}
\label{eq:misalignment_belief_operator_properties}
Given a state space $\States := \SNatures \times \Types$, an agent $i \in \Agents$, and an agent-dependent type structure $\ADTypeSt^i (\Types_i)$: 
\begin{enumerate}[leftmargin=*,label=\arabic*)]
\item (Necessitation) $\ADBel_i (\SNatures \times \ADTypes^{i}_{-i}) = \Types_i$;

\item (Monotonicity) for every events $\ADEvent^{i}_{-i}, \ADEvent^{i'}_{-i}$, if $\ADEvent^{i}_{-i} \subseteq \ADEvent^{i'}_{-i}$, then $\ADBel_i (\ADEvent^{i}_{-i}) \subseteq \ADBel_i (\ADEvent^{i'}_{-i})$;

\item (Conjunction) for every events $\ADEvent^{i}_{-i}, \ADEvent^{i'}_{-i}$, $\ADBel_i (\ADEvent^{i}_{-i}) \cap \ADBel_i (\ADEvent^{i'}_{-i}) = \ADBel_i (\ADEvent^{i}_{-i} \cap \ADEvent^{i'}_{-i})$;

\item (Positive Introspection) for every event $\ADEvent^{i}_{-i}$, $\ADBel_i (\ADBel_i (\ADEvent^{i}_{-i})) = \ADBel_i (\ADEvent^{i}_{-i})$;

\item (Negative Introspection) for every event $\ADEvent^{i}_{-i}$, $\ADBel_i \big( [\ADBel_i (\ADEvent^{i}_{-i})]^\complement \big) = [\ADBel_i (\ADEvent^{i}_{-i})]^\complement$.
\end{enumerate}
\end{proposition}

As pointed out above, the results just stated are conceptually important, because they establish that all the usual properties that are satisfied by the belief operator in the absence of misalignment are still satisfied when we contemplate the possibility of having misalignment.

\section{Misalignment Matters}
\label{sec:misalignment_matters}

We now show that contemplating the presence of misalignment leads to novel behavioral predictions. In particular, given the nature of our framework (i.e., the fact that it is based on opportunely handling CHBs of interest) and the tools employed (i.e., interactive epistemological ones), we make this point building on the seminal \citet[Theorem 1, p.21]{Milgrom_Stokey_1982}, i.e., the famous \Say{No Speculative Trade} theorem, which is built on a state space (implicitly built on CHBs) and on the usage of modal operators.

Recalling that no-trade theorems require a common prior, to formalize this point in our setting we define the events $\evaluation{\snature} := \{\snature\} \times \Types$ and $\evaluation{\type_i} := \SNatures \times \{\type_i\} \times \Types_{-i}$, with $\evaluation{\snature, \type_i} := \evaluation{\snature} \cap \evaluation{\type_i}$. Thus, a \say{standard}  type structure $\TypeSt := \la \Agents, \SNatures , \Rounds { \belfun_j , \Types_j}_{j \in \Agents} \ra$ with corresponding state space $\States := \SNatures \times \Types$ \emph{admits a common prior} if there exists a $\cprior \in \Delta (\States)$ such that, for every $i \in \Agents$ and $\type_i \in \Types_i$:\footnote{To simplify, throughout this section we only consider finite state spaces (as in \Mref{def:state_space}, i.e., state spaces that are not necessarily belief-closed).}
\begin{enumerate}[leftmargin=*,label=\arabic*)]
    \item $\cprior (\evaluation{ \type_i}) > 0$;

    \item for every $\snature \in \SNatures$ and $\type_{-i} \in \Types_{-i}$,
    \begin{equation*}
        \belfun_i (\type_i) \big(\{x, \type_{-i}\}\big) = %
        \cprior \Rounds {\evaluation{x, \type_{-i}} | \evaluation{\type_i}} .
    \end{equation*}
\end{enumerate}
Hence, the famous theorem of \citet[Theorem 1, p.21]{Milgrom_Stokey_1982} can be stated as follows: if we have a \say{standard} type structure that admits a common prior, no-trade is Pareto efficient, and there exists a state in which there is common belief that (another) trade is accepted by every player, then this trade has to be indifferent to no-trade for every player.

Now, we are interested in settings that allow for misalignment, i.e., type structures that are not \say{standard} per \Mref{th:state_space_characterization}, and for this, we need to extend the notion of a common prior too. Thus, we fix a state space $\States = \SNatures \times \Types$, and---given our previous analysis---we contemplate the presence of misalignment via a profile of agent-dependent type structures $\big( \ADTypeSt^j (\Types_j ) \big)_{j \in \Agents}$. One natural extension of the common prior would then be to also consider a corresponding profile of agent-dependent common priors $\big( \cprior^j \big)_{j \in \Agents}$ so that $\ADTypeSt^i (\Types_i)$ admits $\cprior^i$ as a common prior for every $i \in \Agents$. Mathematically---but not conceptually---this is actually equivalent to just allowing heterogeneous priors, but possibly on differing state spaces. We want a further consistency condition that makes the agent-dependent common priors agree on the likelihoods of those states for which there is no misalignment, which we formalize as follows.

Given a state space $\States = \SNatures \times \Types$ and a corresponding pair of profiles $\big( \ADTypeSt^j (\Types_j ), \cprior^j \big)_{j \in \Agents}$, we say that $\cprior \in \Delta(\States)$ is a \emph{consistent prior} if 
\begin{enumerate}[leftmargin=*,label=\arabic*)]
    \item for every $\omega \in \States$, $\cprior (\state) > 0$, and

    \item for every $\omega' \in \States \cap\ADStates^i$
    \begin{equation*}
    \frac{\cprior(\omega)}{\cprior(\omega')} = \frac{\cprior^i(\omega)}{\cprior^i(\omega')}
    \end{equation*}
for every $i \in \Agents$.
\end{enumerate}
Note that the second condition is essentially the usual requirement of Bayesian updating. However, since we do not necessarily have $\States \subseteq \ADStates^i$ nor $\States \supseteq \ADStates^i$, we only require the likelihoods to be preserved on the intersection.\footnote{See \cite{Karni_Viero_2013} for a similar notion declined to account for the presence of unawareness.}

If the profile of agent-dependent type structures $\big( \ADTypeSt^j (\Types_j ) \big)_{j \in \Agents}$ is common, then the existence of a consistent prior immediately renders all agent-dependent common priors equal too, and therefore we obtain an immediate extension of the no-trade theorem to a specific form of misalignment. We state the theorem only informally to avoid introducing all the extraneous notation needed to formalize the notion of trade environment.

\begin{theorem}[Generalized No-Speculative-Trade Theorem]
\label{th:generalized_no-speculative-trade_derived}
Given a finite state space $\States$, a common profile of agent-dependent type structures $\big(\ADTypeSt^j (\Types_j ) \big)_{j \in \Agents}$ with $\ADStates^i$ finite for every $i \in \Agents$, a profile of agent-dependent common priors $\big( \cprior^j \big)_{j \in \Agents}$, and a consistent prior $\cprior$, if no-trade is Pareto efficient and there exists a state in which there is common belief that (another) trade is accepted by every player, then this trade has to be indifferent to no-trade for every player.
\end{theorem}

According to our taxonomy of misalignment, \Mref{th:generalized_no-speculative-trade_derived} does not address whether or not speculative trade is possible if the agent-dependent state spaces are non-common (and therefore, also non-degenerate). Indeed, our running example shows that such a trade is possible.

\begin{example}[continues=ex:misalignment, name=Speculative Trade with a Non-Common \& Non-Degenerate Profile of Agent-Dependent Type Structures]
As before, we have $\Agents := \Sets { \Ann , \Bob }$ and $\SNatures := \Sets { \snature^{\rain} , \snature^{\sun} }$, where the actual state space of interest for us as analysts is 
\begin{equation*}
\States :=  \SNatures \times \{\type^{\rain}_\Ann\} \times \{ \type^{\sun}_\Bob\} \equiv %
\{(\snature^{\rain}, \type^{\rain}_\Ann , \type^{\sun}_\Bob), (\snature^{\sun}, \type^{\rain}_\Ann , \type^{\sun}_\Bob)\} \equiv \Sets { \state^1 , \state^2 } .
\end{equation*}
Recall the minimal agent-dependent type structures from before, for Ann and Bob, respectively:
\begin{align*}
\minADTypeSt^{\Ann} (\Types_\Ann) &:= 
\Angles { \Agents, \SNatures , \Rounds {\minADTypes^{\Ann}_j , \minADbelfun^{\Ann}_j }_{j \in \Agents} } , & \text{ and } &&
\minADTypeSt^{\Bob} (\Types_\Bob) &:= 
\Angles { \Agents, \SNatures , \Rounds {\minADTypes^{\Bob}_j , \minADbelfun^{\Bob}_j }_{j \in \Agents} }
\end{align*}
with  
\begin{align*}
\minADTypes^{\Ann}_{\Ann}  &:= \{ \type^{\rain}_{\Ann} \} ,
& && \minADTypes^{\Bob}_{\Bob} & := \{ \type^{\sun}_{\Bob} \},
\\
\minADTypes^{\Ann}_{\Bob} & := \{ \type^{\rain}_{\Bob} \}, 
&&&\minADTypes^{\Bob}_{\Ann} & := \{ \type^{\sun}_{\Ann} \},
\end{align*}
and
\begin{align*}
\ADbelfun^{\Ann}_{\Ann} (\type^{\rain}_{\Ann}) & := 1 (\snature^\rain, \type^{\rain}_{\Bob}) ,
&&&\ADbelfun^{\Bob}_{\Bob} (\type^{\sun}_{\Bob}) & := 1 (\snature^\sun, \type^{\sun}_{\Ann}),
\\
\ADbelfun^{\Ann}_{\Bob} (\type^{\rain}_{\Bob}) & := 1 (\snature^\rain, \type^{\rain}_{\Ann}) ,
&&& \ADbelfun^{\Bob}_{\Ann} (\type^{\sun}_{\Ann}) & := 1 (\snature^\sun, \type^{\sun}_{\Bob}) ,
\end{align*}
so that $\minADStates^\Ann := \{\state^{\rain}\}$ and $\minADStates^\Bob := \{\state^{\sun}\}$. Clearly, for every $i \in \Agents$, there is a unique $\cprior^i \in \Delta(\minADStates^i)$ such that $\minADStates^i$ admits it as a common prior. Here, we can also find a (non-unique) consistent prior $\cprior \in \Delta(\States)$. For example, setting $\cprior(\state^1) = \cprior(\state^2) = 1/2$ yields one which satisfies the requirements of an original common prior. However, it is easily seen that for the profile of agent-dependent type structures, we can construct a trade that both are willing to take and even have common belief of that in their respective agent-dependent type structures.
\end{example}

The following table summarizes the speculative trade possibilities in accordance with out taxonomy.

\begin{table}[!ht]
\begin{center}
\renewcommand{\arraystretch}{1.2}
\centering
\begin{tabular}{|c|c|c|} 
\hline
					& Common 	& Non-Common \\ 
\hline
\rule{0pt}{3ex}
Degenerate     		&   	\citet{Milgrom_Stokey_1982}			&  $\times$    \\
Non-Degenerate 	&  \Mref{th:generalized_no-speculative-trade_derived} 				&    \Mref{ex:misalignment} %
\\ 
\hline
\end{tabular}
\end{center}
\caption{A taxonomy of the trade scenarios captured by addressing misalignment.}
\label{tab:misalignment_taxonomy_trade}
\end{table}

\section{Discussion}
\label{sec:discussion}

\subsection{Conceptual Aspects}
\label{subsec:conceptual_aspects}

\paragraph{Misspecification} Moving from \cite{Berk_1966}, recent times have seen a blossoming of the literature on \emph{misspecification}, where we have misspecification when an agent rules out the distribution that generates a signal she has to receive. Of course, misalignment is \emph{not} directly related to misspecification, since, whereas the former is a phenomenon that concerns \emph{interactive} beliefs alone, the latter concerns agents' beliefs regarding a parameter space (and distributions on them) alone: in other words, whereas misspecification can arise in presence of one agent only, this is not possible for misalignment. However, it is important to observe that misalignment and misspecification share one point in common, namely, the fact that in both cases we are in presence of agents ruling out \say{true} aspects of the model used to formalize their beliefs: under misspecification, agents rule out the \say{true} signal-generating distribution, whereas under misalignment (in particular when non-common and non-degenerate) they rule out the \say{true} CHBs of the other agents.

\paragraph{Elicitability of Misalignment \& CHBs} Throughout this work we argued that one of the main properties (i.e., Property (3) in \Sref{sec:introduction}) of misalignment is its elicitability. In particular, there is an aspect behind this statement that needs to be uncovered, namely, the potential elicitation process of misalignment and how it affects the \say{form} of the corresponding state space. Indeed, for example, if we decide to elicit various orders of beliefs of subjects in a lab in order to check if there is misalignment in an interaction under scrutiny, what we actually obtain is \emph{one} CHB\footnote{Of course, \emph{actually} only up to a certain belief order.} for every agent, that is, building a state space from the elicitation process, we would get that every type space of every agent would be a singleton. Incidentally, in light of this point, it becomes apparent that misalignment is the rule rather than the exception: indeed, in order \emph{not} to have misalignment, we would need to have that, no matter which agent we focus on, the \emph{only} type that captures the CHB of that agent puts probability $1$ on the \emph{only}---elicited---types (as CHBs) of the other agents.

\paragraph{Misalignment \& Experiments on Games} \say{Standard} type structures have recently been used in \cite{Kneeland_2015} to formalize the notion of \emph{rationality and common belief in rationality}, and to identify higher-order rationality in an experimental setting. A close reading of the key definition in \citet[Section 3]{Kneeland_2015} reveals that, for her purposes, the specifics of the type space are not essential---only the induced CHBs matter. As a result, the possibility of misalignment between participants is immaterial to her analysis. However, there are well-known concepts in game theory that depend not only on the (induced) CHBs but also on the underlying state space.\footnote{Examples in game theory include \emph{Bayes-Nash equilibrium}, \emph{interim rationalizability} \emph{à la} \cite{Ely_Peski_2006}, \emph{rationality and common strong belief in rationality} of \cite{Battigalli_Siniscalchi_2002}, and \emph{rationality and common assumption of rationality} of \cite{Brandenburger_et_al_2008}.} In such cases, the use of \say{standard} type structures is \emph{not} without loss of generality, even when used merely as representations of CHBs. This is particularly relevant in experimental settings where, as argued in the previous paragraph, agents may be misaligned in their reasoning. Our approach, based on agent-dependent type structures, allows for the explicit modeling and analysis of such misalignment between participants.

\paragraph{Common Priors \& Heterogeneous Priors}  The literature on CHBs, state spaces, and No-Trade Theorems has typically been linked to the notion of common prior.\footnote{See, for example, \cite{Morris_1994}, \cite{Samet_1998}, \cite{Bonanno_Nehring_1999}, \cite{Feinberg_2000}, \cite{Heifetz_2006}, \cite{Hellman_Pinter_2024} for various characterizations of the notion of common prior along with the corresponding implications for No-Trade-Theorems.} Now, in this work we take the common prior simply as a technical device that ensures \say{consistency} among the beliefs of different agents.\footnote{See the exchange given by \cite{Gul_1998} and \cite{Aumann_1998b}.} Given this, contemplating the presence of misalignment shows that the notion of common prior as addressed so far in the literature contains implicit assumptions that the analysis in \Sref{sec:misalignment_matters} uncovers with respect to the dichotomy \Say{common vs. heterogeneous priors}. That is, what can be seen as a collection of heterogeneous priors under \say{standard} state spaces could be considered a profile of agent-dependent common priors for common agent-dependent state spaces.

\paragraph{Cognitive Phenomena} Various models have been proposed to deal with cognitive\footnote{See \Sref{foot:cognitive} for a description of how we employ the term \Say{cognitive}.} phenomena such as \emph{information processing mistakes} (as in \cite{Geanakoplos_2021}), \emph{unawareness}\footnote{With respect to this point, see also how the literature on unawareness has dealt with speculative trade, as in \cite{Heifetz_et_al_2006}, \cite{Heifetz_et_al_2013b}, and \cite{Galanis_2018}.} (as in \cite{Heifetz_et_al_2006}) and \emph{framing} (as in \cite{Charness_Sontuoso_2023}). Misalignment is indirectly related to these phenomena in the way in which we capture it, that is, via appropriately modifying the notion of state space of interest to model a given environment with interactive beliefs.

\subsection{Technical Issues}
\label{subsec:technical_issues}

\paragraph{Domain of Uncertainty} In \Sref{sec:background_knowledge}, we started from a compact space $\SNatures$ of states of nature, where this was a choice made for presentation purposes to simplify the exposition in light of the application tackled in \Sref{sec:misalignment_matters}. However, it is understood that we could have phrased everything by simply starting with an \emph{abstract} compact space $X$ acting as a---common---domain of uncertainty, thus, remaining silent on the specific properties of $X$. As a result, by proceeding along these lines, these properties would change according to the application in mind, where we would capture the scenario tackled in the present work with $X := \SNatures$. In particular, this would also allow to capture a situation in which agents can---possibly---have private information by setting $X := (\SNatures_j)_{j \in \Agents_0}$ as the set of \emph{payoff--relevant states}, with $\SNatures_i$ denoting the set of \emph{agent $i$'s payoff--relevant states} and $\SNatures_0$ the set of \emph{states of nature}. Finally, as it is customary in epistemic game theory, $X$ could also capture the uncertainty of agents---as players---concerning the choices of the other agents---as co-players---in a game theoretical environment: in particular, rather than being a set of payoff--relevant states, the abstract space $X$ could be explicitly defined as $X := (S_j)_{j \in \Agents}$, with $S_i$ being the strategy space of an arbitrary agent $i$---considered as a player in a game of interest.

\paragraph{Topological Closure} By inspecting Condition 2(b) in \Mref{def:minimal_agent-dependent_type_structure} and \Mref{eq:agent_closure} we see that both definitions have to rely on additional topological tools to make them well-defined, that is, the addition of types at the (topological) boundary in Condition 2(b) in the definition of minimal agent-dependent type structure and the application of the topological closure operator in the definition of the agent closure. Since the agent closure provides a foundation for the notion of minimal agent-dependent type structure captured in \Mref{def:minimal_agent-dependent_type_structure}, we focus here on the agent closure. Thus, in the present context, the application of the topological closure operator is a technical device that is needed to make the definition of agent closure (as a state space) well-behaved with respect to our definition of belief-closed state space as set forth in \Mref{def:belief-closed_state_space}. However, it has to be pointed out that this does not have an impact on the analysis performed here, since the agent closure is based on the application of the closure operator as defined in \Mref{eq:closure_operator}, which is in itself based on the $\supp$ operator. Indeed, crucially, the $\supp$ operator treated as a correspondence is lower hemicontinuous (as established in \citet[Theorem 17.14, p.563]{Aliprantis_Border_2006}), which has as an immediate consequence the fact that employing the topological closure does not add any \say{new} (unwanted) types beyond those that belong to the boundary. This is particularly relevant since finite \say{standard} type structures are the go-to objects to typically perform analysis and, as a result, the topological closure does not play any role in \emph{finite} minimal agent-dependent type structures induced by the agent closure.

\paragraph{Partitional Information Structures} It is important to observe that agent-dependent type structures are \emph{partitional}\footnote{See \citet[Definition 68.2, p.68]{Osborne_Rubinstein_1994} and \citet[Chapter 9]{Maschler_et_al_2013} (in particular Chapter 9.4).} in nature, which is immediate in light of the fact that they are nothing more than \say{standard} type structures in their own rights with the additional---and crucial---twist that they capture the CHBs of \emph{all} agents that stem from the CHBs of \emph{one} specific agent. Now, this is---among other things---what makes the present framework technically different from the one presented in \cite{Geanakoplos_2021}, where the focus is on non-partitional structures.

\subsection{Further Venues}
\label{subsec:further_venues}

\paragraph{Game Theory} Building on \Sref{subsec:conceptual_aspects} \Say{Misspecification} and \Sref{subsec:technical_issues} \Say{Domain of Uncertainty}, the novel framework introduced here seems particularly apt to study solution concepts based on misspecification, since it provides an additional dimension to the analysis that can be performed when in presence of misspecification \emph{and} multiple interactive players. Thus, with respect to this point, the present framework could be used---among other things---to provide an epistemic\footnote{Concerning epistemic game theory, see \cite{Perea_2012}, \cite{Dekel_Siniscalchi_2015}, and \cite{Battigalli_et_al_Forthcoming}.} foundation of those solution concepts covered under the Berk-Nash Equilibrium of \cite{Esponda_Pouzo_2016}\footnote{See also  \cite{Kalai_Lehrer_1995}.} in the spirit of the epistemic characterization of Nash Equilibrium of \cite{Aumann_Brandenburger_1995}.

\paragraph{Measuring Misalignment} In the spirit of the use of the Kullback-Leibler divergence of \cite{Kullback_Leibler_1951} as it is employed in the literature on misspecification to measure that phenomenon, it could be important to identify and axiomatically characterize a measure of misalignment.

\paragraph{Learning about Misalignment} A natural question that stems from the present work is if agents can learn about the misalignment present in the interactive environment along with the potential behavioral implications arising in a trade environment. As such, this kind of research question would fit in---and extend---the literature stemming from \cite{Geanakoplos_Polemarchakis_1982}.\footnote{See \cite{Gilboa_et_al_2022} for a take on the matter with a focus on \say{large worlds}, i.e., settings that lack an understanding of the underlying uncertainty that is commonly accepted by the agents.}

\paragraph{Contemplating Own Misalignment} In the same spirit in which it is possible for agents to be aware of their own unawareness,\footnote{As discussed, for example, in \citet[Chapter 3.5]{Schipper_2015}.} it could be possible for an agent to be aware of having misaligned beliefs. Thus, the nature of the behavioral predictions that arise in such a scenario should be investigated.


\appendix

\section*{Appendix}

\section{Proofs}
\label{app:proofs}


\subsection{\texorpdfstring{Proofs of \Sref{sec:defining_misalignment}}{Proof of Section 3}}
\label{subapp:proof_defining_misalignment}

It has to be observed that the construction of the space of CHBs can be performed as in \cite{Mertens_Zamir_1985} with an arbitrary agent $i \in \Agents$ having beliefs over herself: when we add---as it is customary---the additional assumption that $\marg_{\UTypes_i} \belhom_i (\type_i) = \delta_{\type_i}$, we end up with the same object we obtained in \Sref{sec:background_knowledge}. Now, for the proof that follows we assume to have performed the construction with agents having beliefs over themselves. In particular, this leads to the definition of the $m^{\text{th}}$-order universal hierarchy function of agent $i \in \Agents$ as 
\begin{equation*}
\label{eq:app_mth_hierarchy_function}
    \hierfun^{m}_i : \UTypes_i \to \Delta (\SNatures \times \Types^{m-1}) ,
\end{equation*}
which is in the spirit of \Mref{eq:hierarchy_function} for $m \geq 2$. 

We can now proceed with the proof of \Mref{th:state_space_characterization}.

\begin{proof}[Proof of \Mref{th:state_space_characterization}]
Note that the statement is logically equivalent to the following: a state space $\States$ is aligned if and only if it is
belief-closed. This is the statement we prove.\\
\noindent $[\Rightarrow]$ Suppose the state space $\States := \SNatures \times \Types$ is aligned. That is, for every agent $i \in \Agents$, every type $\type_i \in \Types_i$, every $m \in \Naturals \setminus \{1\}$, and every agent $j \in \Agents \setminus \{i\}$ such that 
\begin{equation}
\label{eq:alignment}
\Big( \supp \marg_{\Hiers^{m-1}_{j}} \hierfun^{m}_i (\type_i) \Big) \not\subseteq \projstfun^{m-1}_j\Types_j .
\end{equation}
Thus, we can follow exactly the arguments (modulo the differences just mentioned above), of \citet[Proof of Property 5]{Mertens_Zamir_1985} to establish belief-closedness.

\noindent $[\Leftarrow]$ Suppose the state space $\States := \SNatures \times \Types$ is misaligned, but belief-closed. Then, there exist agents $i, j \in \Agents$ with $i \neq j$, $m \in \Naturals \setminus \{1\}$, $\type_i \in \Types_i$, and $\type_j^* \in \UTypes_j \setminus \Types_j$ such that
\begin{equation*}
\projstfun^{m-1}_j\type_j^* \in  \supp \marg_{\Hiers^{m-1}_{j}} \hierfun^{m}_i (\type_i),
\end{equation*}
which is in direct contradiction to belief-closedness.
\end{proof}

\subsection{\texorpdfstring{Proofs of \Sref{sec:a_framework_for_misalignment}}{Proof of Section 4}}
\label{subapp:proof_a_framework_for_misalignment}

In order to prove \Mref{th:characterization_minimality}, we employ order theoretical arguments. For this, we need a pointed complete lattice $\big((L,\leq), \star\big)$  (i.e., $\star \in L$). Recall that a map $F: L \to L$ is \emph{extensive} if $x \leq F (x)$ for every $x \in L$. $F : L \to L$ is \emph{$\omega$-continuous}\footnote{To differentiate the order-theoretic notion of continuity from our notion of continuity in the main text, we use the terminology \Say{$\omega$-continuity} here. We decided to use that expression in line with the terminology employed in category theory to refer to the same notion.} if for any ascending chain $(D_n)_{n \in \mathbb{N}}$, we have
\begin{equation*}
    F \Rounds {\bigvee_{n \in \mathbb{N}} D_n} = \bigvee_{n \in \mathbb{N}} F (D_n).
\end{equation*}

Now, we consider a map $F: L \to L$ that is extensive and $\omega$-continuous map. Set $K_{0} := \star$ and define inductively
\begin{equation*}
K_{n+1} := F(K_{n})
\end{equation*}
for every $n \in \Naturals_0$. Also, let
\begin{align*}
\mathcal{P}_{n}&:=
\left\{X \in L\mid  \star \leq X, \ F^{n}(\star)\leq X\right\},\\ 
A_{n} & :=
\bigwedge_{X \in \mathcal{P}_{n}} X.
\end{align*}
Finally, let
\begin{equation*}
K_{\infty} := \bigvee_{n \in \Naturals_0}K_{n} ,
\end{equation*}
which is well-defined because $L$ is complete.

\begin{lemma} 
The meet $A_{n}$ exists in $L$ with $A_{n}\in \mathcal{P}_{n}$.
\end{lemma}

\begin{proof}
Since $L$ is complete, the meet exists in $L$.  Moreover, each $X\in\mathcal{P}_{n}$ satisfies $\star \leq X$ and $F^{n}(\star)\leq X$, hence, their meet $A_{n}$ also satisfies these two conditions. Thus, $A_{n}\in\mathcal{P}_{n}$.
\end{proof}

\begin{lemma}
    For every $n \in \Naturals_0$, $A_{n}=K_{n}$.
\end{lemma}

\begin{proof}
First, $K_{n}\in\mathcal{P}_{n}$ because $\star \leq K_{n}$ (from the fact that $F$ is extensive) and $F^{n}(\star)=K_{n}\leq K_{n}$. Hence,
\begin{equation*}
A_{n}
=
\bigwedge_{\,X\in\mathcal{P}_{n}}X
\;\leq\; K_{n}.
\end{equation*}
Conversely, for each $X\in\mathcal{P}_{n}$ we have $F^{n}(\star)=K_{n}\leq X$, so $K_{n}$ is a lower bound of $\mathcal{P}_{n}$.  Since $A_n$ is the greatest lower bound $K_{n} \leq A_n$
\end{proof}

\begin{lemma}
    The set $K_{\infty}$ is the least fixed point of $F$.
\end{lemma}

\begin{proof}
Note that $(K_n)_{n \in \mathbb{N}}$ is an ascending chain. Since $F$ is $\omega$-continuous, we have 
\begin{equation*}
    F \Rounds { \bigvee_{n \in \Naturals_0} K_{n} }
=
\bigvee_{n \in \Naturals_0} F \big(K_{n} \big),
\end{equation*}
where the LHS is equal to $F\big(K_{\infty}\big)$ by definition and the RHS is equal to $\bigvee_{n \in \Naturals_0} K_{n+1} = \bigvee_{n \in \Naturals_0} K_{n} = K_{\infty}$. Thus, $F\big(K_{\infty}\big)=K_{\infty}$ 
To establish that $K_{\infty}$ is the least fixed point of $F$ let
\begin{align*}
\mathcal{F}&:=
\big\{X\in L \mid \star\leq X, F(X)=X\bigr\}, \\
A_{\infty}&:=
\bigwedge_{X \in \mathcal{F}} X.
\end{align*}
Since $L$ is complete, the meet $\bigwedge_{X\in\mathcal{F}} X$ exists.  For every $X \in \mathcal{F}$, we have
\begin{equation*}
K_{n} = F^{n}(\star) \leq F^{n}(X) = X .
\end{equation*}
Thus, taking joins over $n$ gives
\begin{equation*}
K_{\infty} = \bigvee_{n \in \Naturals_0}K_{n}\leq X.
\end{equation*}
Hence, $K_{\infty}$ is a lower bound of $\mathcal{F}$, so
\begin{equation*}
K_{\infty} \leq \bigwedge_{X \in \mathcal{F}}X
=
A_{\infty}.
\end{equation*}
On the other hand, $A_{\infty}\in\mathcal{F}$, because $\star \leq A_{\infty}$, and
\begin{equation*}
F\big(A_{\infty}\big)
\leq
\bigwedge_{\,X\in\mathcal{F}}F(X)
=
\bigwedge_{\,X\in\mathcal{F}}X
=
A_{\infty} \leq F\big(A_{\infty}\big),
\end{equation*}
where the last inequality uses the fact that $F$ is extensive.
\end{proof}

We now transpose the result obtained above in an abstract order theoretical framework to our game theoretical setting. Thus, for this purpose we fix a state space $\States$ and we work with compact sets and the lattice induced by set inclusion (component-wise with respect to the cartesian product indexed by $\Agents$). Note that all our sets are subsets of the universal state space $\UStates$, which is compact itself: as a result, we only need to establish the closedness of the subsets to get compactness. However, the lattice with unions as joins is not complete, because only finite unions of closed sets are closed. Thus, we define the following join operation to make our lattice complete:
\begin{equation}
\label{eq:join_operation}
    \bigvee_{\lambda \in \Lambda} D_\lambda := \overline{\bigcup_{\lambda \in \Lambda} D_\lambda},
\end{equation}
for any family of compact sets indexed by $\Lambda$. Let $\CompactFam(\UStates)$ denote the collection of compact subsets of $\UStates$, i.e., $\States := \SNatures \times \prod_{j \in \Agents} \Types_j \in \CompactFam (\UStates)$ if $\Types_i \subseteq \UTypes_i$ is compact for every $i \in \Agents$.

\begin{lemma}
\label{lem:join_operation}
The tuple  $(\CompactFam(\UStates), \subseteq)$ with join operation as defined in \Mref{eq:join_operation} and meet given by intersection is a complete lattice.
\end{lemma}

\begin{proof}
Let $(\States_{\lambda})_{\lambda\in\Lambda}\subseteq \CompactFam(\UStates)$ be any family of compact sets.  Now, their union  $U = \bigcup_{\lambda\in\Lambda} \States_{\lambda}$ is not necessarily closed, but it is contained in the compact space $\UStates$. Let
\begin{equation*}
\bigvee_{\lambda\in\Lambda} \States_{\lambda}
:=
\overline{\,U\,}
=
\overline{\bigcup_{\lambda\in\Lambda} \States_{\lambda}}.
\end{equation*}
Since $U\subseteq \UStates$ and $\UStates$ is compact, $\overline{U}$ is closed in $\UStates$ and hence compact.  Clearly for every $\States_{\lambda}$ we have $\States_{\lambda} \subseteq U \subseteq \overline{U}$. Hence, $\overline{U}$ is an upper bound.  If any compact $Y\subseteq \UStates$ satisfies $\States_{\lambda}\subseteq Y$ for every $\lambda \in \Lambda$, then $U\subseteq Y$, and $Y$ being closed forces $\overline{U}\subseteq Y$.  Therefore
\begin{equation*}
\overline{\bigcup_{\lambda\in\Lambda} \States_{\lambda}}
\end{equation*}
is the least upper bound of $(\States_{\lambda})$ in $\CompactFam(\UStates)$.  The infimum case is automatically satisfied, because arbitrary intersections are closed. 
\end{proof} 

Now, to apply our order-theoretic setting we also need the pointed element $\star$ in the lattice and the mapping $F$. Thus, for this we fix an arbitrary $i \in \Agents$, we set
\begin{equation*}
    \star := \overline{\States \cup \Closure_i (\States)}
\end{equation*}
(which does depend on the arbitrary agent $i$), and we let
\begin{equation}
\label{eq:extensive_map}
    F\left(\States'\right) := \States' \cup \bigcup_{j \in \Agents}\Closure_j (\States') 
\end{equation}
for every $\States' \in \CompactFam(\UStates)$, which is by construction an element in $\CompactFam(\UStates)$, i.e., it is compact. Furthermore, it is immediate that $F$ is extensive. Thus, we only need to establish that $F$ is also $\omega$-continuous.

\begin{lemma}
The map $F$ as defined in \Mref{eq:extensive_map} is $\omega$‐continuous in $\CompactFam(\UStates)$.
\end{lemma}

\begin{proof}
Let $(\States^{n})_{n \in \Naturals_0}$ be an ascending chain in $\CompactFam(\UStates)$, where
\begin{align*}
    \States^{n} = 
\SNatures \times \prod_{j\in\Agents} \Types_{j}^{n},
\end{align*}
and $\Types_{i}^{n}\subseteq \Types_{i}^{n+1}$ for every $i \in \Agents$ and $n \in \Naturals_0$.  Thus, by our definition of the join, we have
\begin{align*}
\bigvee_{n \in \Naturals_0} \States^{n}
=
\SNatures  \times 
\prod_{j\in\Agents}\overline{\bigcup_{n \in \Naturals_0} \Types_{j}^{n}}
=
\SNatures  \times 
\prod_{j\in\Agents}\bigvee_{n \in \Naturals_0} \Types_{j}^{n}.
\end{align*}
Letting $\Types_i := \bigvee_{n \in \Naturals_0} \Types_{i}^{n}$ for every $i \in \Agents$, we have
\begin{align*}
    \Closure_i \left(\bigvee_{n \in \Naturals_0}\States^{n} \right) =
    \SNatures \times \Types_i \times \overline{\supp \typehom_i (\Types_i)}
\end{align*}
and
\begin{align*}
    F\left(\bigvee_{n \in \Naturals_0}\States^{n} \right)=\bigvee_{n \in \Naturals_0}\States^{n} \cup \bigcup_{j \in \Agents} \left(\SNatures \times \Types_j \times \overline{\supp \typehom_j (\Types_j)}\right).
\end{align*}
Furthermore,
\begin{align*}
    F\left(\States^{n} \right)=\States^{n} \cup \bigcup_{j \in \Agents} \left(\SNatures \times \Types_j^{n} \times \overline{\supp \typehom_j (\Types_j^{n})}\right).
\end{align*}
Thus, we only need to show that 
\begin{equation*}
    \overline{\supp \typehom_i (\Types_i)} = \bigvee_{n \in \Naturals_0} \overline{\supp \typehom_i (\Types_i^{n})} = \overline{\bigcup_{n \in \Naturals_0} \overline{\supp \typehom_i (\Types_i^{n})}} 
\end{equation*}
for every $i \in \Agents$. Thus, let $i \in \Agents$ be arbitrary. First, we establish 
\begin{equation*}
    \overline{\bigcup_{n \in \Naturals_0} \overline{\supp \typehom_i (\Types_i^{n})}} \subseteq \overline{\supp \typehom_i (\Types_i)} .
\end{equation*}
Since $\Types_{i}^{n}\subseteq \Types_i$, we have $\supp\typehom_{i} (\Types_{i}^{n}) \subseteq \supp\typehom_{i} (\Types_i)$ and also $\overline{\supp\typehom_{i}(\Types_{i}^{n})} \subseteq \overline{\supp\typehom_{i} (\Types_i)}$. Hence
\begin{align*}
    \bigcup_{n \in \Naturals_0}  \overline{\supp\typehom_{i} (\Types_{i}^{n})} \subseteq \overline{\supp\typehom_{i}(\Types_i)}.
\end{align*}
Taking the closure (together with their idempotence) gives the result.

\noindent We now prove the reverse inclusion
\begin{equation*}
    \overline{\supp \typehom_i (\Types_i)} \subseteq %
    \overline{\bigcup_{n \in \Naturals_0} \overline{\supp \typehom_i (\Types_i^{n})}} .
\end{equation*}
Thus, consider $\type_{-i} \in \supp\typehom_{i} (T_i)$. From the fact that the $\supp$ operator is lower hemicontinuous and from the continuity of $\typehom_{i}$, there exists a sequence $t_i^{m} \in \Types_i$ with $t_i^{m} \to t_i \in \Types_i$ and $\type_{-i}^{m} \in \supp \typehom_{i} (t_i^{m})$ such that $\type_{-i}^{m} \to \type_{-i}$. By the property of closure, we can (by introducing more sequences if needed) make sure that each $t_i^{m} \in \bigcup_{n \in \Naturals_0} \Types_i^{n}$. With this, each $t^{m}_i$ lies in some $\Types_{i}^{n_{m}}$. Because we consider an increasing sequence of sets, we can assume that $n_1 \leq n_2 \leq \ldots$ without loss. Thus, for every $m \in \Naturals$,
\begin{align*}
    \type_{-i}^{m} \in \supp\typehom_{i} (\Types_{i}^{n_m}) \subseteq \overline{\supp\typehom_{i} (\Types_{i}^{n_m})}.
\end{align*}
As $\type_{-i}^{m} \to \type_{-i}$, we conclude that
\begin{align*}
    \type_{-i} \in \overline{\bigcup_{n \in \Naturals_0} \overline{\supp \typehom_i (\Types_i^{n})}}.
\end{align*}
Since $\type_{-i}$ was arbitrary, we actually have
\begin{align*}
    \supp \typehom_{i} (\Types_i) \subseteq \overline{\bigcup_{n \in \Naturals_0} \overline{\supp \typehom_i(\Types_i^{n})}}.
\end{align*}
Taking the closure (together with their idempotence) gives the result again.
\end{proof}

\clearpage
\phantomsection\addcontentsline{toc}{section}{References}
\hypersetup{colorlinks=true,linkcolor=green!50!black}

\end{document}